\documentclass[fleqn,usenatbib]{mnras}

\usepackage{newtxtext,newtxmath}

\usepackage[T1]{fontenc}
\usepackage{soul}

\DeclareRobustCommand{\VAN}[3]{#2}
\let\VANthebibliography\thebibliography
\def\thebibliography{\DeclareRobustCommand{\VAN}[3]{##3}\VANthebibliography}
\usepackage{graphicx}
\usepackage{amsmath}
\usepackage{hyperref}
\usepackage{xcolor}
\usepackage{makecell}
\usepackage{multirow}

\title[Radio Strong Lensing]{Strong gravitational lensing with upcoming wide-field radio surveys}

\author[McCarty et al.]{
Samuel McCarty$^{1, 2, 3}$,\thanks{E-mail: smmccrty@uw.edu \& liam.connor@cfa.harvard.edu}
Liam Connor$^{2}$
\\
$^{1}$Department of Astronomy, University of Washington, Seattle, WA 98195-1580, USA\\
$^{2}$Center for Astrophysics | Harvard \& Smithsonian, Cambridge, MA 02138-1516, USA\\
$^{3}$Cahill Center for Astronomy and Astrophysics, MC\,249-17, California Institute of Technology, Pasadena CA 91125, USA
}

\date{Accepted 2025 August 15. Received 2025 August 14; in original form 2024 November 01.}

\pubyear{\the\year{}}

\begin{document}
\label{firstpage}
\pagerange{\pageref{firstpage}--\pageref{lastpage}}
    \maketitle

\begin{abstract}
The number of strong lensing systems will soon increase by 
orders of magnitude thanks to sensitive, wide-field optical and infrared imaging surveys such as Euclid, Rubin-LSST, and Roman. 
A dramatic increase in strong lenses will also occur at
radio wavelengths. The 2000-antenna Deep Synoptic Array (DSA-2000) will detect $\sim10^9$ continuum
sources in the Northern Hemisphere with a high mean redshift ($\langle z_s \rangle \approx2$), the Square Kilometre Array mid frequency telescope (SKA-Mid) will observe a large sample of extragalactic sources in the South with sub-arcsecond resolution, and the Very Large Array Sky Survey (VLASS) has recently completed.
We forecast lensing rates for these telescopes, finding that each of the DSA-2000 and SKA-Mid will conservatively discover $\mathcal{O}(10^4)$ strongly lensed systems, and optimistically as many as $\mathcal{O}(10^5)$, a significant fraction of which will be galaxy group and cluster lenses. We propose strategies for strong lensing discovery in the limit where the Einstein radii are comparable to the PSF angular scale, taking advantage of modern computer vision techniques and multi-survey data.  Finally, we describe applications of the expected radio strong lensing systems, including
time-delay cosmography with transient and variable sources. 
We find that $\sim$30-300 time-variable flat-spectrum AGN discovered by the DSA-2000 and SKA-Mid 
could be used to constrain $H_0$ at the percent level
with the appropriate follow-up.
\end{abstract}




\section{Introduction}
Strong gravitational lensing has a multitude of applications in astrophysics and cosmology \citep{Treu2010}. Previously theoretical ideas have been put into practice in recent decades as the number of known lensed systems has increased and observational data have improved. For example, strongly lensed time-variable and transient sources can be 
used to constrain the Hubble constant, $H_0$, because the time-delay of a multiply imaged source depends 
on the geometry of the Universe \citep{refsdal}. The technique is known as time-delay cosmography.
With just six lensed quasar systems, The $H_0$ Lenses in COSMOGRAIL’s Wellspring
(H0LiCOW) 
collaboration has reported 2.4\,$\%$ precision on their $H_0$ measurement, which is independent of the distance ladder and the CMB \citep{Wong_2019}.

In addition to the Universe's large-scale geometry, lensing 
observables are sensitive to the total mass of the deflector 
galaxy or cluster, allowing one to measure the spatial 
distribution of matter and test different dark matter models \citep{massey_2010, Vegetti2024}.
Lensing magnification allows astronomers to 
observe objects in the distant Universe, as was pointed out at the field's inception \citep{zwicky1937}. 
Dramatic examples have come from the James Webb Space Telescope (JWST), including a red supergiant star at 
$z\approx2.2$ that appears to be magnified by a factor of several thousand due to its proximity to caustics in a cluster lens \citep{jwstclusterlens}. 

Nearly all of these applications 
would benefit from a larger sample of strong lensing 
systems. To date, order 
$10^3$ confirmed strong lensing systems have been discovered, most of 
which were identified at optical and infrared wavelengths (O/IR)\footnote{\url{https://sled.amnh.org/}}. Fortunately, 
upcoming wide-field imaging surveys such as Euclid 
and The Vera C. Rubin Observatory's Legacy Survey of Space and Time (Rubin-LSST)
are each expected to detect as many as $\sim$\,10$^5$ strong lenses 
\citep{Collett_2015}. Early releases from Euclid have recently affirmed these forecasts \citep{euclid2024,euclidcollaboration2025euclidquickdatarelease}.
\cite{Weiner_2020} predict 
the Nancy Grace Roman Space Telescope's 2000 square degree survey could find of order 20,000 strong lenses, while a recent forecast predicts 160,000 \citep{Wedig2025}. Many more could be found if the proposed multi-epoch $4\pi$\,sr survey 
is carried out \citep{JesseRoman2023}. An increase in the total number of lenses by two-orders of 
magnitude will usher in a new era of strong lensing science. 

The first strongly lensed system ever discovered was co-detected at radio wavelengths \citep{walsh1979}. The first survey for lenses, the MIT-Green Bank survey, was also in the radio \citep{MITGreenBankSurvey}. Some of the first lens time-delays were measured for radio lenses \citep{vanOmmen1995,Lovell1998,Fassnacht1999,Haarsma1999,Biggs1999,Koopmans2000,Patnaik2001}. Radio observations were essential for early gravitational lensing studies because interferometry, e.g. from the Very Large Array (VLA) \citep{thompson1980very}, provided the resolution necessary to identify lenses, and many of the observable radio sources were radio-loud AGN at high redshift, increasing the probability of lensing. This was exploited in the Jodrell-VLA Astrometric Survey (JVAS) \citep{Patnaik1992}, PMN-NVSS Extragalactic Lens Survey (PANELS), \citep{Winn2000b,Winn2001search}, and the Cosmic Lens All Sky Survey (CLASS) \citep{CLASSI,CLASSII}, which are to date the most complete radio lens surveys.  Sources were filtered first by catalog-level requirements (e.g. separate components that have similar spectral indices) before being visually inspected to confirm lensing. Despite this initial work in the radio, to date only $\sim$\,130 lensing systems have been detected at GHz frequencies  \citep{walsh1979,Lawrence1984,Pramesh1988,Hewitt1988,Langston1989,Kayser1990,Burke1990,Lehar1993,Lehar1993b,Falco1996,Schechter1998,Ibata1999,winn2000pmn,Lehar2001,Winn_2001,Winn2002c,Winn2002b,Lacy2002,CLASSII,Carilli2003,Garrett2005,Haarsma2005,Berciano2007,Boyce2007,Inada_2006,Wucknitz2008,Ghosh2009,Ivison2010,McKean2011,Mckean2011b,Jackson2011,Valtchanov2011,Messias2014,Geach2015,Jackson2015,DessaugesZavadsky2017,Blecher2019,Hartley2021,Mangat2021,Glikman2023,Giulietti_2023,Gross2023,Dux2023,Chen2024,jackson2024radio,martinez2024findinglensedradiosources,dobie2024gaia}. Many of these were discovered in the early 2000s with the JVAS/CLASS surveys, but there have also been significant recent discoveries by targeting O/IR selected lenses with the VLA \citep{jackson2024radio,martinez2024findinglensedradiosources,dobie2024gaia}. There are a handful of large-separation radio cluster scale lenses, many of which were sub-mm selected \citep{Garrett2005,Inada_2006,Berciano2007,Ghosh2009,McKean2011,Jackson2011,DessaugesZavadsky2017}. The discrepancy between the number of known lenses in the O/IR and in the radio is due to the relatively small 
total number of known radio sources ($\sim$\,$10^7$) \citep{Condon_1998,Helfand_2015,Gordon_2020,Duchesne2024racs} and the lack of wide-field radio imaging surveys with $\sim$\,arcsecond resolution. Both of those limitations will soon be overcome with the advent of next-generation radio survey telescopes. 

The 2000-antenna Deep Synoptic Array (DSA-2000) will detect over one billion 
radio sources with a deep redshift distribution, most of which will be star-forming radio galaxies (SFRG) or active galactic nuclei (AGN)  \citep{hallinan2019dsa2000radiosurvey}. The DSA-2000 is expected to see first light in 2027 with key surveys running between 2028 and 2033. Its point-spread function (PSF) will 
be roughly $2''$ at the top of the 0.7--2\,GHz radio band. A 
fifty-fold increase in the total radio source catalog is made possible by the DSA-2000's high survey speed, driven by a large field-of-view ($\sim$\,10\,deg$^2$) and high sensitivity (the expected  system-equivalent flux-density or "SEFD" is just 2.5\,Jy). Its cadenced all-sky survey will map out the $3\pi$\,sr above declination 
$-30^\circ$ down to 500\,nJy/beam root-mean square noise \citep{hallinan2019dsa2000radiosurvey}. While the nominal cadence 
of the continuum survey is four months, certain fields 
may be visited more regularly, which will enable the identification of more lensed time-variable sources and possibly present an opportunity to measure 
lensing time-delays with better temporal sampling.

The full mid-frequency telescope for the Square Kilometre Array (SKA-Mid AA4) will consist of 197 fully steerable 13.5\,m dishes (including the existing MeerKAT radio telescope), operating between 350\,MHz and 15.4\,GHz with a field-of-view of roughly 1\,deg$^2$ at 1400\,MHz \citep{braun2019anticipatedperformancesquarekilometre}. It is unclear when the full AA4 configuration will be completed, but construction of an intermediate configuration, the AA* of SKA-Mid, with 144 dishes is anticipated to end in 2031\footnote{\url{https://www.skao.int/en/explore/telescopes/ska-mid}}.
Although the SKA-Mid's lower survey speed will result in 
fewer sources than the DSA-2000, the 
wide frequency range and long maximum 
baseline (150\,km for AA4, 40\,km for AA*) will enable high-resolution imaging, an asset to strong lensing studies \citep{mckean2015ska}. In the longer term, the Next Generation Very Large Array (ngVLA) is a planned interferometer with 
extraordinary sensitivity covering a wide range of frequencies (1.2--116\,GHz) \citep{ngVLAspie2}. It will be able to resolve features at
milliarcseconds scales. While its broad science goals 
did not require optimizing the instrument 
for mapping speed \citep{ngVLAspie}, the ngVLA will be a 
world-class instrument for strong lensing science.

Radio strong lensing offers distinct advantages to studies at shorter wavelengths and will complement O/IR imaging surveys. Arguably the most important of these advantages is the extremely high angular resolution that can be achieved with Very Long Baseline Interferometry (VLBI), which is crucial for precision lens modeling (e.g. \cite{Spingola2019}, \cite{Powell2021}, and \cite{Stacey2024}). SFRGs and radio AGN can be detected to great distances, boosting the mean optical depth of radio continuum sources \citep{Saxena2018,Gloudemans_2022}. Obscuration 
by dust in lensing galaxies is not an issue at radio wavelengths, 
nor is the variable "seeing" that impacts ground-based O/IR telescopes. Relatedly, the point-spread function (PSF) of a radio interferometer is directly determined by observing frequency and array configuration, allowing us to accurately forward model the instrument's response. In the limit of a 
large number of antennas and a filled aperture (a "radio camera"), the deterministic PSF allows us to 
be more ambitious in image-plane deconvolution, enabling techniques such as super-resolution \citep{POLISH}. Radio telescopes also measure full 
polarization information, 
which is conserved under gravitational lensing \citep{Greenfield85, Dyer1992}. 
Finally, the larger emission regions of radio AGN render them less 
susceptible to microlensing and may provide cleaner modeling of the deflector mass distribution \citep{Birrer2024}. 
This is critical for measuring $H_0$ via time-delay cosmography.

Radio lensing is key for three of the main applications of strong lensing: time-delay cosmography, dark matter studies, and source science. Polarization information can be used to measure time-delays between multiple images for $H_0$, in some cases allowing better constraints on the time-delay than photometry alone \citep{Biggs1999,Biggs2018b,Biggs2018,Biggs2021,Biggs2023}. One of the six H0LiCOW lenses is a radio selected lens \citep{Wong_2019,Fassnacht1999,Fassnacht_2002b}. The high angular resolution of interferometry is instrumental for the detection of sub-galactic dark matter halos at cosmological distances through gravitational lensing \citep{Dalal_2002,Vegetti_2012,MacLeod_2013,Hsueh2016,Hsueh2018,Hsueh2020}. The rotation measure (RM) of a lensed source allows 
one to measure the magnetic field and ionized gas properties of the lensing galaxy and intervening 
halos at cosmological distances \citep{Mao_2017}. All of these applications will benefit from a larger sample of known radio lenses.

There are of course several drawbacks to strong 
lensing studies in radio surveys, often precluding the use of 
standalone radio observations. 
For example, continuum 
sources will not contain redshift information, and 
multi-wavelength datasets or follow-up will be necessary 
for the majority of lensed systems at radio frequencies. Secondly, the morphology of radio lobes can mimic lensing arcs, exacerbating the already-challenging lens identification problem (consider, for example, head-tail galaxies 3C\,465 or 3C\,129 at redshift 2). In the case of the DSA-2000, SKA-Mid AA* (at least at 1.4\,GHz), and VLASS, the angular resolution is such that the majority of galaxy-galaxy lensing systems will be unresolved in the absence of super-resolution \citep{Oguri2006}. 

In this work we seek to study radio strong lensing in upcoming interferometric surveys, and develop strategies that  leverage the unique capabilities of radio lensing while alleviating its drawbacks. 
We first forecast strong lensing 
rates for different source and deflector classes for the SKA-Mid, VLASS, and for the first time the DSA-2000. We compare our results with previous radio lensing forecasts and forecasts for next generation O/IR surveys. Next, due to the arcsecond scale PSFs of these telescopes, we discuss methods for increasing the efficiency of lens searches in the radio, including super-resolving lensing candidates with modern computer vision techniques, in order to increase the yield of strongly lensed systems. We then consider lensed time-variable 
and transient sources that these radio surveys will find and 
discuss their utility for measuring $H_0$
with the appropriate follow-up, as well as other applications of strong lensing. 

\section{Methods}
\label{sec:model}

\subsection{Lens Model}
\label{sec:lensmodel}

We seek to develop a realistic model of deflectors and sources to predict the number of strongly lensed systems expected in upcoming radio surveys. Previous lensing forecasts for O/IR surveys have typically focused on galaxy-scale lenses, which are modeled as Singular Isothermal Ellipsoids \citep{Oguri2010,Collett_2015,LensQSOsim}. Galaxy group and cluster scale lenses are neglected because they make up a smaller, but still significant, fraction of the total lenses. Additionally, lens modeling is more complicated in this regime.  From \cite{Oguri2006}, the number of group and cluster lenses are around 11\% and 3\% of the total galaxy lenses, respectively (excluding sub-halo lensing). However, these systems make up a considerable portion of lens systems with angular separation of order 1'', and almost all of the lenses at $\geq$10''. Because the resolutions of the radio telescopes considered in this investigation are relatively low (Section \ref{sec:dlimit}), these lenses will be a considerable fraction of the discoverable lensing systems and so must be accounted for in our investigation. We define galaxies, galaxy groups, and galaxy clusters as systems with mass $M_\text{h}<10^{13}\,M_\odot$, $10^{13}\,M_{\odot}\leq M_\text{h}<10^{14}\,M_{\odot}$, and $M_\text{h}\geq10^{14}\,M_{\odot}$ respectively, where $M_\text{h}$ is the dark matter halo mass. In reality, the galaxy scale lenses are limited to $M_\text{h}\gtrsim10^{11}\,M_\odot$ because systems below this mass will have lensing separations below the pixel resolution scale of our simulations, and the cluster scale lenses are limited to $M_\text{h}\lesssim10^{16}\,M_\odot$ because systems above this mass are exceedingly rare. 

To model the population of deflectors we use the halo-based approach of \cite{Oguri2006, Abe_2025}. This model is desirable for our purposes because it produces a smooth transition between the galaxy, galaxy group, and galaxy cluster regimes while being very computationally efficient due to algorithms developed by \cite{Oguri_2021}. It provides a well-motivated model for groups and clusters and has been verified to reproduce the population of deflectors observed in galaxy-scale lensing effectively. The code for calculating lens parameters is adapted from the SL-Hammocks\footnote{\url{https://github.com/LSSTDESC/SL-Hammocks}} code \citep{Abe_2025}. We include an abbreviated description of the model below; see \cite{Abe_2025} for a full description and validation. 

Each deflector is modeled as a dark matter halo with a central elliptical galaxy stellar component. The population of halos is determined by the halo mass function from \cite{Tinker_2008}. For a given halo mass $M_\text{h}$, the density profile of the dark matter component is set as a Navarro-Fenk-White profile (NFW) \citep{NFW}. The concentration parameter $c$ of the profile is determined from the mass-concentration relation of \cite{Diemer_2015}, with updated parameters from \cite{Diemer_2019}, and includes a lognormal scatter of $\sigma_{\text{ln}c}=0.33$. The halo mass function and halo parameters are computed with the \texttt{COLOSSUS} package \citep{Diemer_2018}. Each NFW halo is modified with a line-of-sight projected ellipticity $e_\text{h}$. The ellipticity is computed as a truncated normal distribution with a mean ellipticity determined by the mass as in \cite{Okabe_2020}, a standard deviation of $\sigma_{e_{\text{h}}}=0.13$, and truncation at 0.0 and 0.9. The halo is also given a position angle counterclockwise from the y-axis in the lens plane $\phi_\text{h}$ from a uniform distribution between -180 and 180 degrees. 

Each halo is assigned a central galaxy whose mass is computed using the stellar mass-halo mass relation of \cite{Behroozi_2019}, modified for the Salpeter IMF \citep{Salpeter} by \cite{Abe_2025}. A lognormal scatter is included in this relationship with $\sigma_{\text{ln}M_*}=0.2$. The stellar profile is set as the Hernquist profile \citep{Hernquist}. The mean effective radius for the Hernquist profile $r_{\text{e}}$ is computed from a fitting function employed by \cite{Abe_2025} and fit to JWST early-type galaxy data from \cite{vanderwel2023stellarhalfmassradii05z23} including lognormal scatter. The stellar ellipticity $e_*$ follows a lognormal distribution with mean 0.3, $\sigma_{\text{ln}e_*}=0.16$, and truncation at 0.0 and 0.9. The galaxy position angle $\phi_*$ is correlated to $\phi_h$ using a gaussian distribution centered on $\phi_h$ with $\sigma_{\phi_*}=35.4$\,deg. 

Finally, an external shear $\gamma_{\text{ext}}$ is included to account for line-of-sight effects. The shear has two components with a lognormal scatter that evolves with redshift as in \cite{Abe_2025}. We do not account for sub-halo lensing or external convergence. It is shown in \cite{Oguri2006,Abe_2025} that a significant fraction of lenses, roughly half for groups and clusters and 10\% for galaxies, come from lensing by sub-halos. The environment around these sub-halos boosts their lens capabilities and image separations. Further, it is well known that intervening masses along the line of sight and deflector environments can increase the probability of multiple imaging by a significant fraction, especially at high source redshifts and large image separations, which we also neglect in this study \citep{Oguri_2005, Fleury_2021}. Several of the CLASS lenses have been shown to reside in groups and many of the lenses have line-of-sight structures that affect the lensing potential \citep{Fassnacht_2002,McKean2005,Momcheva_2015}. This means that we may be underestimating the number of lenses considerably, especially group and cluster scale lenses. Because these higher-order aspects of the model are likely to \textit{increase} the number of lenses, for the purpose of this paper it is safe to ignore them. However, it should be noted that these effects will alter the image separation distribution of the lens systems.

\subsection{Source Populations}
\label{sec:sourcepops}

\begin{figure*}
	\includegraphics[width=0.8\textwidth]{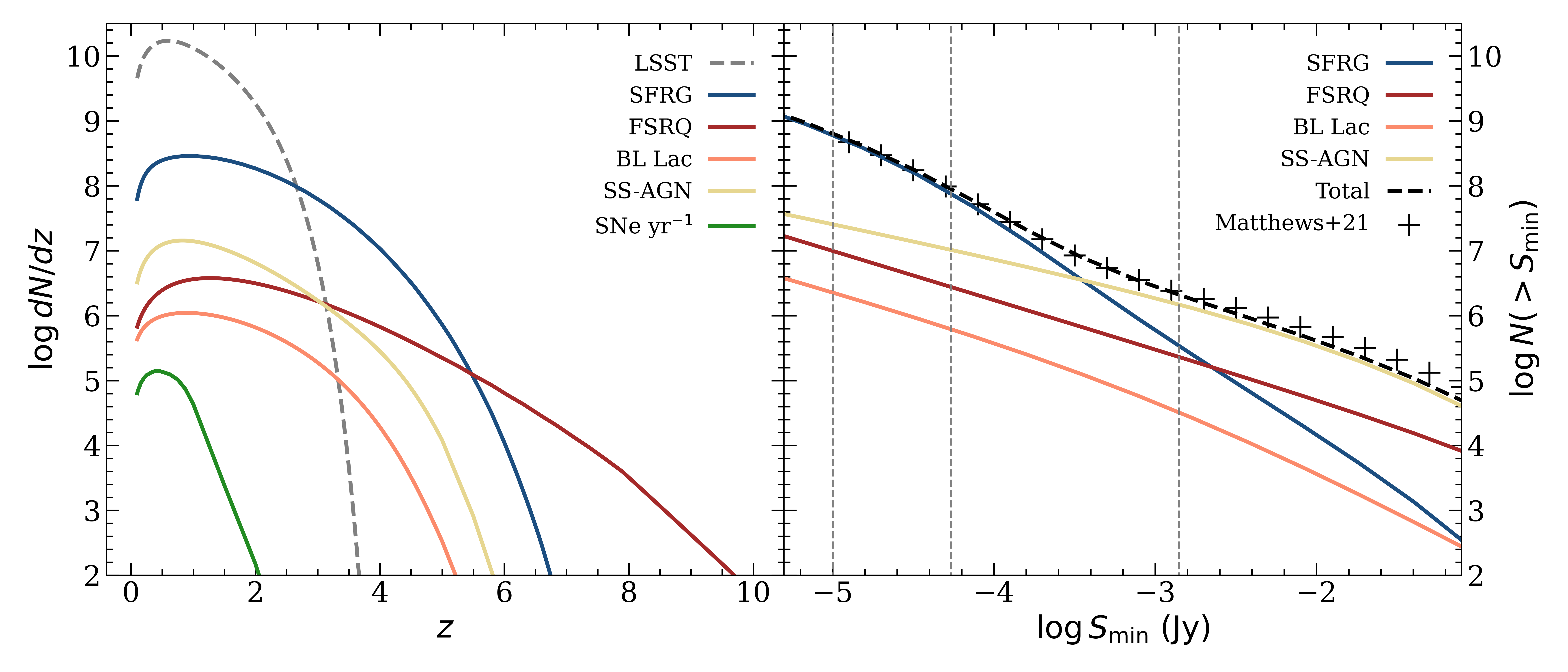}
    \caption{Left: expected source redshift distribution above the DSA-2000 point source flux-density limit (10$\,\mu$Jy) from the model of Section \ref{sec:sourcepops}. The distribution for Rubin-LSST is taken from \protect\cite{Alonso_2015}. Right: expected cumulative source counts above a minimum flux-density $S_\text{min}$ compared to the observationally determined source counts from \protect\cite{Matthews_2021}. The vertical dashed grey lines are the adopted $S_\text{min}$ for the DSA-2000, SKA-Mid AA*, and VLASS from left to right.}
    \label{fig:source}
\end{figure*}
 
To get redshift distributions for different source classes, we model the Star-Forming Radio Galaxy (SFRG) and radio Active Galactic Nucleus (AGN) populations using the luminosity functions (LF) from the Tiered Radio Extra-galactic Continuum Simulation (T-RECS) \citep{TRECS}. These have been shown to match observations out to high redshift. We ignore the dust contribution to the SFRG luminosity, which is not significant at $\lesssim$100\,GHz \citep{TRECS}. For the star formation rate (SFR) function, we use the analytic fit from \cite{Mancuso_2015}. Following \cite{TRECS}, we divide the AGN into three sub populations: Flat Spectrum Radio Quasars (FSRQ), BL-Laceratae (BLLac), and Steep Spectrum AGN (SS-AGN). At 1.4\,GHz, AGN dominate the source population at $S\gtrsim1$\,mJy, while SFRGs dominate below \citep{TRECS}.

From the LF we calculate the number of sources in a redshift interval as
\begin{equation}
\label{eqn:nlim}
    N(>L_\text{min},z) =  dz \int_{L_{\text{min}}(z)} d\text{log}L \,\Phi(L, z)  \,\frac{d^2V_c}{d\Omega dz_d} \,\Omega_{\text{survey}}
\end{equation}
\noindent where $\frac{d^2V}{dz_d d\Omega}=(1+z_d)^3 \frac{c\,dt}{dz_d} D_d^2$ is the differential comoving volume with $D_d$ the angular diameter distance, $\Omega_{\text{survey}}$ is the sky coverage of the survey in steradians ($\approx3\pi$ for the surveys considered in this work) and $\Phi (L,z)$ is the LF of the source population. 
\begin{equation}
    \label{lmin}
    L_{\text{min}}(z)=\frac{4 \pi D_L^2}{(1+z)^{1+\alpha}} \left(\frac{\nu_e}{\nu_o} \right)^\alpha S_{\text{min}}
\end{equation} 
\noindent is the minimum observable luminosity at a redshift $z$, where $D_L$ is the luminosity distance, $\alpha$ is the spectral index of the source population, $S_{\text{min}}$ is the minimum detectable flux-density for the survey (see Section \ref{sec:dlimit}), $\nu_e$ and $\nu_o$ are the emitted and observed frequencies, and the factor $(1+z)^{-(1+\alpha)}$ is the standard cosmological radio K-correction. 

In addition to SFRGs and AGN, we build a model for radio SNe. To date, only $\sim$100 radio core-collapse supernovae (ccSNe) have been detected and the first Type Ia radio supernova was detected recently \citep{Bietenholz_2021, Kool2023}. To model the expected rate of radio ccSNe we follow \cite{Lien_2011}. Because ccSNe are short-lived, the rate of ccSNe is closely related to the cosmic SFR. For the radio luminosity distribution of the ccSNe, we use that of \cite{Bietenholz_2021} because it incorporates all known radio detections and non-detections of SNe. 
The radio ccSNe observations used in \cite{Bietenholz_2021} have frequencies ranging from 4-10\,GHz, we assume the LF rest frame frequency is 6\,GHz and $\alpha=-0.7$ for the synchrotron emission of radio ccSNe.
Further, none of the known radio ccSNe have luminosities above 10$^{29}\,\text{erg}\,\text{s}^{-1}\,\text{Hz}^{-1}$, but given the wide spread we expect that the highest luminosities could be several times larger. Considering this, we set an exponential cutoff in linear space on the lognormal distribution of radio ccSNe luminosities at 10$^{30}\,\text{erg}\,\text{s}^{-1}\,\text{Hz}^{-1}$ to avoid including a small number of unphysical ccSNe in our model. We also include a restriction on ccSNe detectability based on their light curves and survey cadence, described in the next section. We scale the number of ccSNe predicted by our model for VLASS to the empirically determined rate (a factor of order 1, see Section \ref{sec:results}). We do not include other radio transients, such as tidal disruption events (TDEs), gamma ray burst (GRB) afterglows, or fast radio bursts (FRBs), either because a lack of observational data makes them difficult to forward model or predictions for their lensing rates already exist (a simple empirical estimate for lensed TDE rates is given in Section \ref{sec:results}; an FRB forecast can be found in \citet{connorfrblensing}).

The predicted redshift distributions of SFRGs, radio AGN, and radio ccSNe above the DSA-2000 point source flux-density limit ($10\,\mu$Jy, Section \ref{sec:dlimit}) are shown on the left in Figure \ref{fig:source} and compared to the expected distribution of Rubin-LSST sources. As expected, the total source count is dominated by SFRGs until high redshift. Blazars, meaning both FSRQs and BLLacs, are the AGN that will be most useful for time-delay measurements, which is explored in Section \ref{sec:h0}. These are the most common sources at very high redshift. We expect radio telescopes to probe higher redshifts than the Rubin-LSST because radio emission is much less susceptible to intervening gas or seeing. This is one of the reasons that the DSA-2000 and SKA-Mid will be effective at lens finding. The right plot in Figure \ref{fig:source} shows the cumulative source counts predicted by our model compared to the observationally determined source counts from \cite{Matthews_2021}.

\subsection{Discoverability limit and expected survey performances}
\label{sec:dlimit}

\begin{figure}
	\includegraphics[width=\columnwidth]{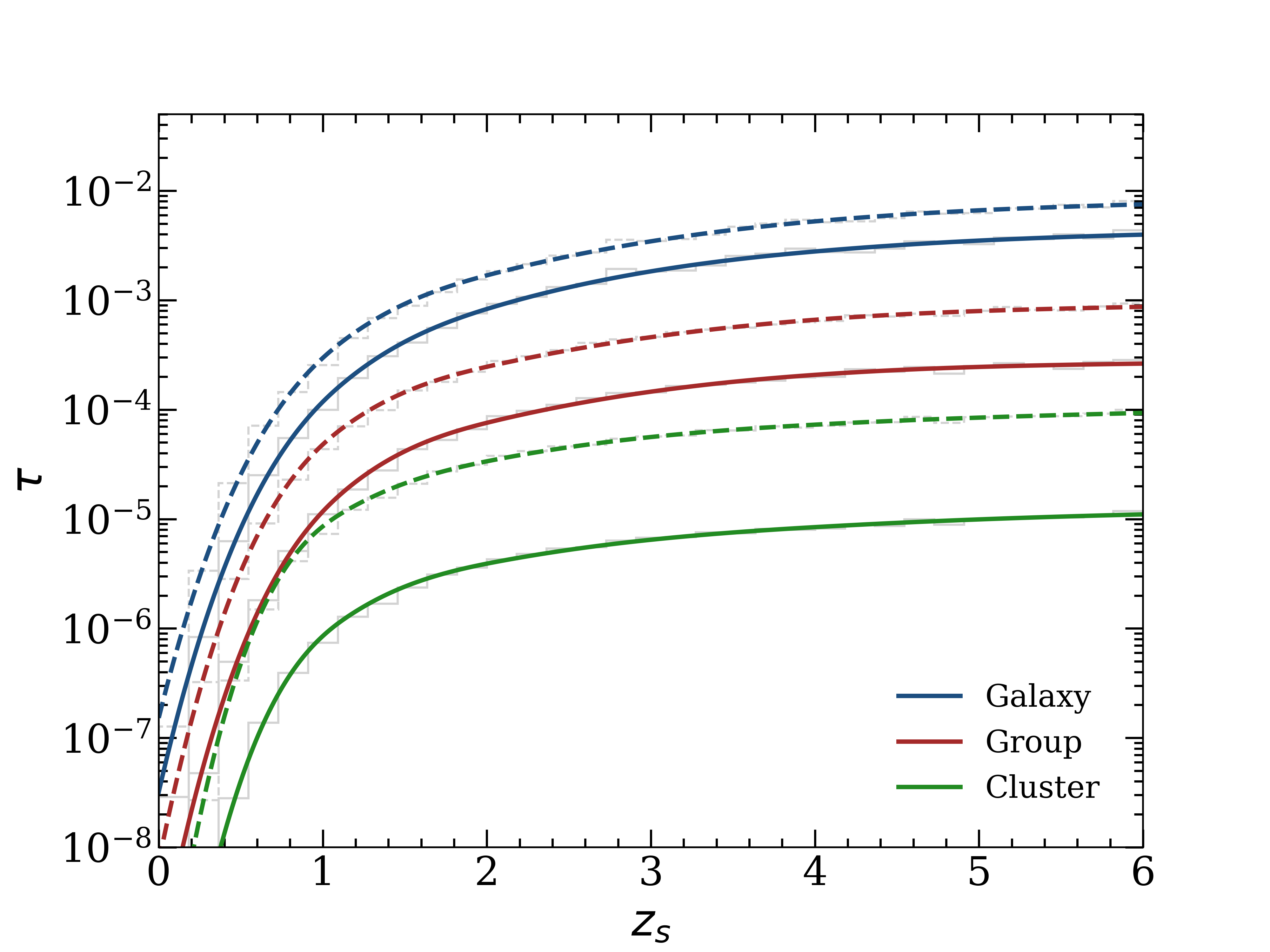}
    \caption{Lensing optical depth vs. redshift, with (dashed) and without (solid) the magnification bias for SFRGs, for a flux-density limit of $10\,\mu$Jy (DSA-2000). The curves are smooth interpolations overlaid on the discrete distributions calculated from our simulation in grey.}
    \label{fig:tau}
\end{figure}

The quantity of interest is not the number of strong lenses contained in a survey but the number of lenses that could be identified. The main parameters determining the discoverability of a lens are its maximum image separation $\Delta\theta$ (when compared to the resolution of the survey instrument) and the signal-to-noise ratio (SNR) of the lensed images, although there are many other effects on the selection function (e.g. the flux-density of the deflector) \citep{Collett_2015,euclidcollaboration2025euclidquickdatarelease}. O/IR lensing forecasts for quasars typically require $\Delta\theta\geq2/3\times\theta_\text{PSF}$ \citep{Oguri2010,LensQSOsim}, where $\theta_\text{PSF}$ is the FWHM of the PSF of the telescope.  
\cite{Abe_2025} do not include a constraint on $\Delta\theta$ because there has been significant work on detecting unresolved lensed quasars and SNe. \cite{Collett_2015} requires $\Delta\theta\geq\theta_\text{PSF}$ in the case of a point source, while \cite{Wedig2025} require $\Delta\theta\geq3\times\theta_\text{PSF}$ for the $0.1''$ resolution of Roman. For the SNR, \cite{Oguri2010,LensQSOsim,Abe_2025} require that more than one image be above the SNR detection threshold (10, 5, and 10 $\times$ survey noise respectively), because in general point source quasars/SNe will not be stretched into arcs. \cite{Collett_2015,Wedig2025} require that the total SNR of the combined multiple images is $\text{SNR}_{\text{tot}}\geq20$. \cite{Weiner_2020,Holloway_Verma_Marshall_More_Tecza_2023} adopt the same requirements as \cite{Collett_2015}.

The only study investigating the capability of modern lens finding algorithms such as Convolutional Neural Networks (CNNs) at radio wavelengths is \cite{Rezaei2022}. 
\cite{Rezaei2022} train CNN architectures to identify galaxy-scale lenses in simulated International LOFAR Telescope (ILT) data. They find that $\Delta\theta\geq3\times\theta_\text{PSF}$ and $\text{SNR}_{\text{tot}}\geq20$ is required for reliable detection. Group and cluster-scale lenses are not incorporated in these simulations. Nevertheless, given that this is the only such work, we apply the same criteria for our "conservative" estimate across all lens types (the limit on $\Delta\theta$ is in general less important for group and cluster scale lenses because of their wide separations).

Motivated by several factors, we also offer an "optimistic" estimate that has discoverability limits  $\Delta\theta\geq\theta_\text{PSF}$ and $\text{SNR}_{\text{tot}}\geq20$. Primary among these factors are emerging super-resolution techniques, which show promise for enabling lens identification in the radio at smaller $\Delta\theta$'s for a given $\theta_\text{PSF}$ \citep{POLISH}. Relatedly, the DSA-2000's filled aperture makes the telescope more akin to an O/IR instrument than to LOFAR, meaning that discoverability limits closer to those found in O/IR forecasts may be more appropriate than those of \cite{Rezaei2022} \citep{hallinan2019dsa2000radiosurvey}. Further, combining radio data with data from other wavelengths can increase the efficiency of lens searches, e.g. \cite{martinez2024findinglensedradiosources} identify unresolved lensed quasars in VLASS by cross-matching with O/IR lenses. Given the $\sim$10$^5$ lenses expected in upcoming O/IR surveys \citep{Collett_2015}, this is promising for radio lens finding. Finally, and more speculatively, future ML models trained specifically for the upcoming radio surveys investigated in this work (instead of the ILT) can likely be more effective, given more realistic radio source models, better forward modeling of the instrument response, and advances in ML lens finders. See Section \ref{sec:discover} for a more in-depth discussion of prospects for radio lens finding. 

In this work, we consider the lens finding potential of three radio surveys: the planned DSA-2000 all-sky survey, an SKA-Mid all-sky survey, and the completed VLASS. The DSA-2000 will map $\sim$30,000\,$\text{deg}^2$ of the sky to a combined $\sigma_n=500$\,nJy/beam rms noise \citep{hallinan2019dsa2000radiosurvey}. The average resolution over its 0.7-2\,GHz radio band will be $3''$, while the resolution at the top of the band will be $2''$. In our calculations, we use $\theta_\text{PSF}=2''$ for the DSA-2000 because many of the lenses will be discoverable at the top of the band. The VLASS's complete data products are expected to have $\sigma_n = 70\,\mu$Jy/beam combined and  $\theta_\text{PSF}=2.5''$ for a $\sim$30,000\,$\text{deg}^2$ survey footprint 
\citep{Lacy_2020}. VLASS operates at 3\,GHz, so we correct calculations of Eqn \ref{eqn:nlim} accordingly.

There is currently no concrete plan for an all-sky survey with the SKA-Mid\footnote{See \url{https://www.skao.int/en} for updates.\label{fn:ska}}, but here we consider how well it might do for lens finding with a $30,000\,\text{deg}^2$ survey footprint. Further, it is now unclear when the full AA4 configuration of the SKA-Mid, with 197 dishes and a maximum baseline $\sim$150\,km, will be completed\footnote{\url{https://www.skao.int/en/science-users/599/scientific-timeline}}. An intermediate AA* configuration, with 144 dishes and a maximum baseline of $\sim$40\,km, is expected to be completed by 2031. We perform our calculations for both configurations. We adopt the 1\,hr continuum sensitivity of each configuration as the $\sigma_n$ of the all-sky survey. For AA4 at 1.4\,GHz, this is 
$\sigma_n = 2\,\mu$Jy/beam, and $\theta_\text{PSF}$ will be $0.4''$ \citep{braun2019anticipatedperformancesquarekilometre}. For AA*, $\sigma_n = 2.7\,\mu$Jy/beam and $\theta_\text{PSF}=1.3''$\footnote{\url{https://www.skao.int/sites/default/files/documents/SKAO-TEL-0000818-V2_SKA1_Science_Performance.pdf}}. Our SKA-Mid lensing predictions differ from a previous analysis by \cite{mckean2015ska} because we consider the more realistic short-term AA* configuration, updated AA4 maximum baseline (150km instead of 180km), stricter discoverability limits (\cite{mckean2015ska} require $\Delta\theta>0.3''$),  
group and cluster scale lenses, and a wider range of source classes including time-variable/transient sources.  

The above discussion applies to persistent sources; transients are more complicated. First, because radio ccSNe have a mean rise time of log$_{10}(t_{\text{rise}})=1.7$ days \citep{Bietenholz_2021}, we do not, in general, expect them to be visible in multiple epochs of the surveys considered in this work. To account for this, we use the single epoch sensitivity of each survey to set $S_\text{min}$, which for the DSA-2000 and VLASS are $\sigma_n=2\,\mu$Jy/beam and $\sigma_n=120\,\mu$Jy/beam respectively \citep{hallinan2019dsa2000radiosurvey,Lacy_2020}. For the SKA-Mid, we assume that each field is visited only once in its all-sky survey so that the single and combined epoch sensitivities are the same. Second, we must account for missed SNe due to the time between visits to each field in a cadenced survey. For a given time that a SNe is above the flux-density limit of the survey, $t_\text{vis}$, and a time between observations $C$, the probability that it is captured in the survey is approximately

\begin{equation}
\label{eqn:p}
    p=\text{min}\{t_\text{vis}/C, 1\}\:.
\end{equation}

\noindent $C$ for the DSA-2000 all-sky survey and VLASS are roughly 4 months and 2 years, respectively \citep{hallinan2019dsa2000radiosurvey, Lacy_2020}. While again there is no set plan for an SKA-Mid all-sky survey strategy, in an idealized scenario assuming that each field of the 30,000 deg$^2$ footprint is visited once for 1 hr, and an FOV of 1 deg \citep{braun2019anticipatedperformancesquarekilometre}, the completion time for the survey would be about 3.4 years. Radio ccSNe lightcurves in \cite{Bietenholz_2021} are modeled with an exponential rise and power law decay. Because a given radio ccSNe spends most of its time in the decay, we can use $L(t)\propto(t/t_\text{peak} )^{-\beta}$, with $\beta=1.5$ as in \cite{Bietenholz_2021}, to approximate $t_\text{vis}$ as

\begin{equation}
    t_\text{vis}= \text{max}\left\{t_\text{peak}\left[\left(\frac{L_\text{peak}}{L_\text{min}(z)}\right)^{2/3} - 1\right],\: 0\right\}\:.
\end{equation}  

\noindent Where the distributions of peak luminosities $L_\text{peak}$ and time until peak luminosity $t_\text{peak}$ are the lognormal best fits from \cite{Bietenholz_2021}. By averaging Eqn \ref{eqn:p} over the population of radio ccSNe that have $L_\text{peak}\geq L_\text{min}(z)$, we correct for our use of the distribution of $L_\text{peak}$ to calculate the number of radio ccSNe at a given source redshift (which is not a sufficient constraint for them to be detected) and account for the cadence of the survey. The number of observable radio ccSNe now falls off more quickly with increasing redshift because there are both fewer ccSNe above the flux-density limit and the ccSNe stay above the flux-density limit for less time. It is worth noting that the lensed SNe forecasts from \cite{Oguri2010,Abe_2025} simply require the flux-density from the SNe to be a fixed factor larger than the flux-density limit of the survey. Following \cite{Oguri2010,Abe_2025}, we ignore the bias introduced by the delay between arrival times of a multiply imaged transient (the "time-delay bias" \citep{Oguri_2003}). The number of lensed radio ccSNe expected for each survey per year is multiplied by the total survey duration $t_\text{survey}$ to get the final estimate. 

The discoverability parameters that we use for each radio array are summarized in Table \ref{tab:limits}. The expected number of sources of each type above the $S_\text{min}$ of each survey are shown on the right in Figure \ref{fig:source}.

\begin{table}
    \centering
    \begin{tabular}{lccccc}
    Telescope & \thead{$S_\text{min}$} & \thead{$S_\text{min}$ (S.E.)} & \thead{$\Delta\theta_\text{min}$ (C, O)}  & \thead{$C$} & $t_\text{survey}$ \\
    \hline \hline
    DSA-2000 & 10 & 40 & 6.0, 2.0  & 0.33 & 5.0 \\
    \hline
    \thead{SKA-Mid AA*} & 54 & 54 & 3.9, 1.3 & 3.40 & 3.4\\
    \hline
    \thead{SKA-Mid AA4} & 40 & 40 & 1.2, 0.4 & 3.40 & 3.4 \\
    \hline
    VLA & 1,400 & 2,400 & 7.5, 2.5 & 2.00 & 6.0 \\
    \hline
    \end{tabular}
    \caption{Summary of lens discoverability parameters for the radio telescopes considered in this work. $S_\text{min}$ is the minimum total flux-density from the combined lensed images in $\mu$Jy, and S.E. denotes the single epoch $S_\text{min}$ used for time-variable sources. $\Delta\theta_\text{min}$ is the minimum angular separation between lensed images in arcsec, for our conservative (C) and optimistic (O) estimates. $C$ is the time in years between observations of a field in each cadenced survey. $t_\text{survey}$ is the length of the survey in years.}
    \label{tab:limits}
\end{table}

\begin{figure}
\includegraphics[trim=0cm 0cm 0cm 0cm, clip, width=0.9\columnwidth]{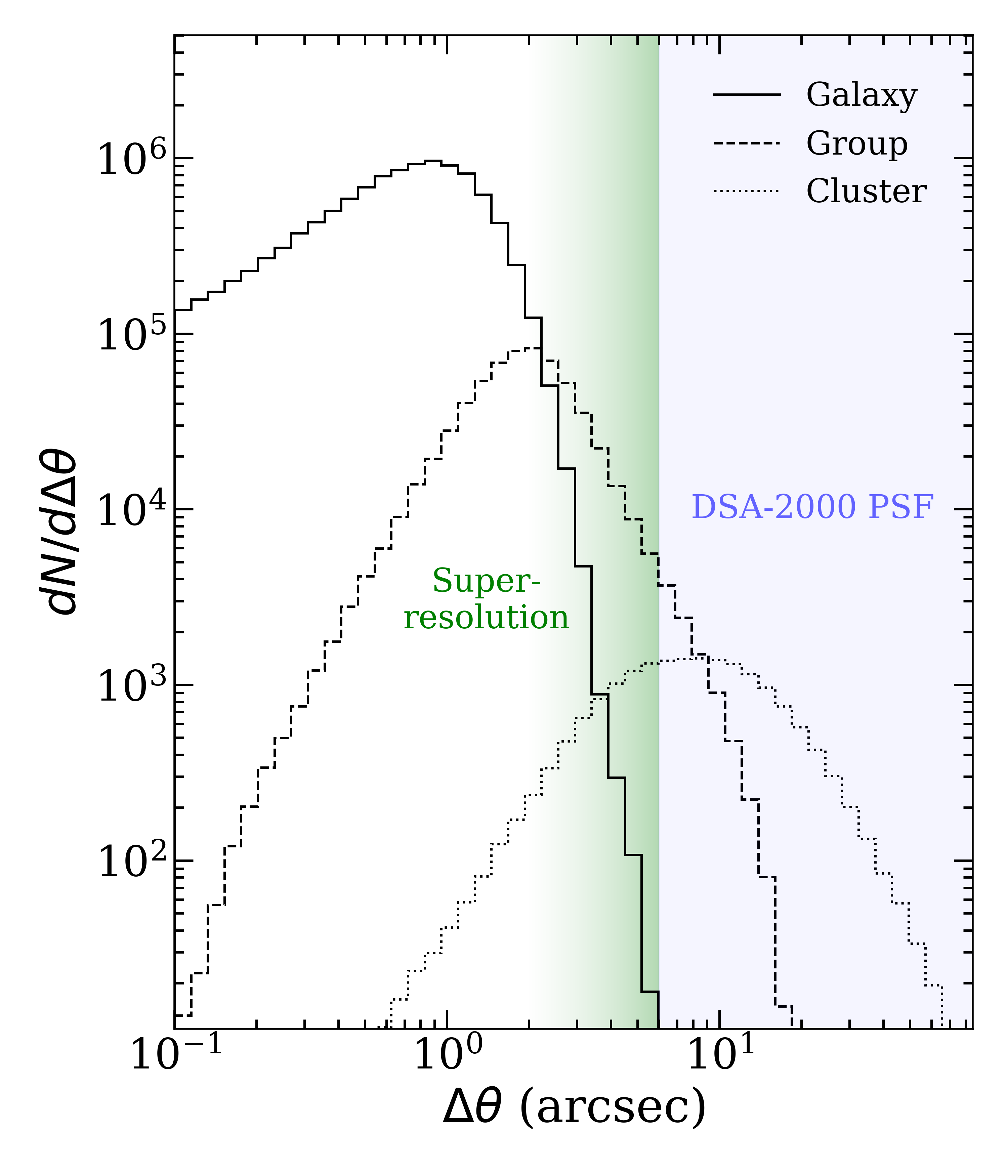}
    \caption{The theoretical image separation distribution of sources strongly lensed by each type of deflector -- galaxies (solid), 
    galaxy groups (dashed), and galaxy clusters (dotted) -- calculated from our model assuming a DSA-2000 source redshift distribution and with magnification bias. The portion that will be discoverable by the DSA-2000 is shown in the blue-shaded region, while the portion that could be discoverable with super-resolution is shown in the green-shaded region.}
    \label{fig:angular_sep}
\end{figure}

\subsection{Optical depth calculations}

For a given source redshift $z_s$, the optical depth for lensing is

\begin{equation}
\label{eqn:tau}
    \tau(z_s) = \int_{0}^{z_s} dz_d \frac{d^2V}{dz_d d\Omega}\int_{M_\text{h min}}^{M_\text{h max}} dM_\text{h} \frac{dn}{dM_\text{h}}\sigma_{\text{lens}} \Theta(\Delta\theta-\Delta\theta_{\text{min}})
\end{equation}

\noindent where $z_d$ is the deflector redshift, $M_\text{h}$ is the mass of the deflector halo, 
$\frac{dn}{dM_\text{h}}$ is the mass function of the deflectors, $M_\text{h min}/M_\text{h max}$ are the mass range limits for the deflector class (galaxy/group/cluster), and $\Theta$ is the Heaviside step function \citep{Oguri2006,Yue_2022}. $\sigma_{\text{lens}}$ is the biased cross-section for multiple imaging by the lens, which we compute as \citep{Huterer2005}:

\begin{equation}
\label{eqn:sigmalens}
    \sigma_\text{lens}(z_s)=\int d\pmb{u} \frac{N(>L_{\text{min}}/\mu(\pmb{u}),z_s)}{N(>L_{\text{min}},z_s)}\,\Theta(n_\text{i}-n_{\text{i min}})
\end{equation}

\noindent where $N(>L_{\text{min}})$  and $L_\text{min}$ are defined in Eqns. \ref{eqn:nlim} and \ref{lmin} respectively, $\pmb{u}$ is the position in the source plane, $\mu(\pmb{u})$ is the magnification at position $\pmb{u}$, and $n_\text{i}$ is the number of multiple images produced by a source at position $\pmb{u}$. 
Note that $\sigma_\text{lens}$ is source dependent due to the LF that appears in Eqn. \ref{eqn:nlim}, so we compute the $\sigma_\text{lens}$ for each source class for each lens. We also calculate $\sigma_\text{lens}$ for $n_\text{i min}=2$, all strong lensing, and $n_\text{i min}=4$, systems which will have extra value due to more constraints on the time-delay or lensing potential. Finally, the total number of lenses expected in a survey is 

\begin{equation}
    \label{eqn:nlens}N_\text{lens}=\int_0^{z_{s\,\text{max}}} dz_s \tau (z_s)\frac{dN}{dz_s}\:.
\end{equation}

In practice, we compute the integral in Eqn \ref{eqn:nlens} with a nested Monte Carlo approach using \texttt{GLAFIC} to perform all lensing calculations \cite{Oguri_2010glafic,Oguri_2021}. The integral of a function $f(\pmb{x})$ can be estimated by 

\begin{equation}
\label{eqn:monte}
    \int f(\pmb{x})\,d\pmb{x} \approx \frac{1}{N} \sum_{i=1}^{N}\frac{f({\pmb{x}_i})}{p(\pmb{x}_i)}
\end{equation}

\noindent where $\pmb{x}$ is a sampled point in the parameter space, $N$ is the number of samples, and $p(\pmb{x}_i)$ is the probability density function used for the sampling evaluated at the $i$th sample \citep{newman1999monte}. First, we draw the independent parameters $M_h$, $z_d$, and $z_l$ from input distributions designed to densely sample important regions while simultaneously covering the full parameter range. The other parameters depend on $M_h$, $z_d$, and $z_l$, and their true distributions in our model are known (Section \ref{sec:lensmodel}). In \texttt{GLAFIC}, we initialize the lens model with the drawn parameters and calculate the location of the caustics for that lens. We then place point sources randomly in the source plane within a region slightly larger than the outer caustic. The number of multiple images produced by each point source and their magnifications are used to estimate Eqn \ref{eqn:sigmalens} using Eqn \ref{eqn:monte}. Note that while we expect many radio sources to be extended, we do not consider extended sources in our lensing calculations. We are, in principle, only attempting to estimate the area for multiple imaging in the source plane, which does not depend on source size; however, extended sources will have complex lensing morphologies and magnifications that will impact their discoverability. This process is repeated for $\sim100,000$ sampled lenses for each survey. The average of $\sigma_\text{lens}$ multiplied by the factors in Eqn \ref{eqn:tau}, 
the number of sources at each redshift $z_s$ (Eqn \ref{eqn:nlim}), and $1/p(\pmb{x}_i)$ from the input distributions gives the final approximation of Eqn \ref{eqn:nlens} via Eqn \ref{eqn:monte}. We can also compute differential distributions of the lens population, e.g. the image separation distribution, by binning the sampled lenses. The number of lenses per deflector class is obtained by ignoring deflectors with halo masses outside the ranges defined in Section \ref{sec:lensmodel}, equivalent to setting $M_\text{h min}/M_\text{h max}$ in Eqn \ref{eqn:tau}. The discoverability limits of Section \ref{sec:dlimit} are enforced in the following way: we ignore the contribution of point sources that produce multiple images with a maximum image separation smaller than $\Delta\theta_\text{min}$ when calculating Eqn \ref{eqn:sigmalens}, i.e. we set $f(\pmb{x}_i)=0$ in Eqn \ref{eqn:monte}, and we use the $S_\text{min}$'s of Table \ref{tab:limits} in all calculations of Eqn \ref{eqn:nlim}. This ensures all lenses satisfy  $\text{SNR}_\text{tot}\geq20$ because the total magnification is always $>1$ (the magnification bias included in Eqn \ref{eqn:sigmalens} automatically adjusts for lensed sources that are intrinsiclly fainter than $S_\text{min}$ but magnified above this limit).

\begin{table*}
\centering
\renewcommand{\arraystretch}{1.4}
\begin{tabular}{l|lc|lc|lc}
    \hline
     & \multicolumn{2}{l|}{SFRG} & \multicolumn{2}{l|}{AGN} & \multicolumn{2}{l}{ccSNe} \\
    Survey & Total & Lensed (C,O) & Total & Lensed (C,O) & Total & Lensed (C,O) \\
    \hline
    \hline
    DSA-2000 & 6.1e8 & 3.1e4 (45\%),\:  1.9e5 (22\%) & 3.8e7
    & 6.9e2 (45\%),\: 5.7e3 (17\%) & 5.3e5
    & 17 (48\%),\: 64 (44\%) \\
    SKA-Mid AA* & 7.7e7 & 1.1e4 (48\%),\: 6.1e4 (27\%) & 1.4e7
    & 3.9e2 (40\%),\: 5.2e3 (9\%) & 7.7e4
    & 4 (51\%),\: 18 (45\%) \\
    SKA-Mid AA4 & 1.2e8 & 1.1e5 (23\%),\: 2.3e5 (20\%) & 1.6e7
    & 7.6e3 (9\%),\: 1.7e4 (8\%) & 1.2e5
    & 31 (40\%),\: 70 (37\%) \\
    VLASS & 1.4e5 & 1.9e1 (54\%),\: 6.5e1 (68\%) & 1.2e6
    & 1.4e1 (44\%),\: 7.3e1 (27\%) & 2.2e3
    & 0 (00\%),\: 0 (00\%) \\
    \hline
\end{tabular}
\caption{Final results of our lensing forecast by source type. "Total" is the total number of sources above $S_\text{min}$. For the number of lensed sources, we show our conservative (C) and optimistic (O) estimates, in that order, as defined in Section \ref{sec:dlimit}. In parentheses are the percentages of lens systems with 4 or more images.}
\label{tab:resultsbysource}
\end{table*}

\begin{table*}
\centering
\renewcommand{\arraystretch}{1.4}
\begin{tabular}{l@{\hspace{1cm}}c@{\hspace{1cm}}c@{\hspace{1cm}}c}
    \hline
     Survey & Galaxy lens & Group lens & Cluster lens \\
    \hline
    \hline
    DSA-2000 & 4.0e0 (100\%),\:  5.1e4 (8\%) & 1.0e4 (55\%),\: 1.2e5 (25\%) & 2.1e4 (42\%),\: 2.5e4 (40\%) \\
    SKA-Mid AA* & 2.5e1 (39\%),\: 3.2e4 (17\%) & 5.5e3 (53\%),\: 2.8e4 (32\%) & 6.1e3 (43\%),\: 6.5e3 (42\%) \\
    SKA-Mid AA4 & 6.5e4 (15\%),\: 2.0e5 (16\%) & 4.1e4 (30\%),\: 4.5e4 (29\%) & 8.4e3 (42\%),\: 8.4e3 (42\%) \\
    VLASS & 0.0e0 (0\%),\: 1.2e1 (35\%) & 4.0e0 (73\%),\: 8.3e1 (49\%) & 2.8e1 (46\%),\: 4.3e1 (45\%) \\
    \hline
\end{tabular}
\caption{Final results of our lensing forecast by deflector type, indicating the number of sources lensed by each deflector type. For the number of lensed sources, we show our conservative  and optimistic estimates, in that order, as defined in Section \ref{sec:dlimit}. In parentheses are the percentages of lens systems with 4 or more images.}
\label{tab:resultsbydeflector}
\end{table*}

\section{Results}
\label{sec:results}

First, we offer a simple empirical forecast based on observational data to compare to our simulation results.

CLASS \citep{CLASSI} sought to study a statistical sample of radio-loud gravitationally lensed systems. We can use CLASS for a rough estimate of the number of lenses the DSA-2000 will find. CLASS found that their statistical sample of 8,958 flat-spectrum radio point sources brighter than 30\,mJy at 5\,GHz had a mean lensing optical depth of $1.5_{-0.3}^{+0.5}\times10^{-3}$ \citep{CLASSII}. By targeting compact sources with flat spectra, many of their objects are AGN at high redshifts (the average source redshift of CLASS lenses is $\sim2$ \citep{CLASSII}), comparable to typical redshifts of DSA-discovered sources, despite the difference in flux scales. Of the confirmed radio lenses in the CLASS statistical sample, 2/13 had angular separations larger than 2 arcseconds, which could be resolved by the DSA-2000 (corresponding to the optimistic estimate from our simulation). Thus, a rough estimate indicates that for every 10,000 sources detected by the DSA-2000, several could be identified as strong lenses. The number of extragalactic sources expected in the DSA-2000's all-sky survey is roughly $\sim10^9$ \citep{hallinan2019dsa2000radiosurvey}. This would result in $\mathcal{O}(10^5)$ new galaxy-scale radio lenses, depending on the practical signal-to-noise threshold for candidate systems. 

We also make an empirical estimate of the rate of lensed radio transients in the DSA-2000 because there is significant uncertainty in forward models for these objects due to the lack of observational data. Lensed transients are also of particular interest for $H_0$ measurements.  We can use data from the Very Large Array Sky Survey (VLASS), which is currently the most comparable survey to the DSA-2000, for a simple estimate \citep{Lacy_2020}. The determined log rates of supernovae (SNe) and tidal disruption events (TDEs) in deg$^{-2}$yr$^{-1}$ are -1.91$^{+0.15}_{-0.16}$ and -2.85$^{+0.28}_{-0.38}$ respectively above 0.7\,mJy (Dong et al. 2025 in prep). If we assume that the total number of observable sources scales as $S_{\text{min}}^{-1.5}$  and the same CLASS optical depth and resolved fraction, the DSA-2000 should find $\mathcal{O}(10)$ galaxy-scale lensed ccSNe and several lensed TDEs during its 5 year survey. 
Further, taking the volumetric rate of optically selected TDEs from \cite{Yao_2023} and assuming 50\% 
emit in the radio at luminosities $\sim10^{37}-10^{39}$\,erg s$^{-1}$ \citep{cendes2023ubiquitouslateradioemission} also indicates several lensed TDEs. So we can tentatively expect to see $\mathcal{O}(1)$ lensed TDEs in the DSA-2000, but a more detailed forward model is needed to be certain. Applying a similar estimate of the lensed rate of GRB afterglows using the predictions of \cite{Ghirlanda2013, Ghirlanda_Burlon_Ghisellini_Salvaterra_Bernardini_Campana_Covino_D’Avanzo_D’Elia_Melandri_etal_2014} gives a rate much less than 1 per year; they are ignored in this investigation.

The DSA-2000 will also discover tens of thousands of distant 
fast radio bursts (FRBs) \citep{petroff, cordes}, some of which will be strongly lensed \citep{connorfrblensing}. 
The key advantage to using FRBs for time-delay lensing is 
that their short duration and coherence allows for exceptional
precision on the gravitational lensing time-delay 
\citep{Wucknitz21}. 
Radio telescopes can preserve phase information about the 
electromagnetic waveform at nanosecond sampling, 
which means microlensing signals can be searched for 
at ultrashort timescales \citep{leung23, kader}. 
However, for cosmological lensing time-delays longer than a pointing 
(i.e. deflectors more massive than $\sim$\,10$^8$\,M$_\odot$),
one needs to catch the lensed images by pointing at the same 
patch of sky when it arrives.
We do not forecast lensed FRB rates here and point the 
reader to previous estimates \citep{connorfrblensing}.

\begin{figure}
    \centering
    \includegraphics[width=0.9\columnwidth]{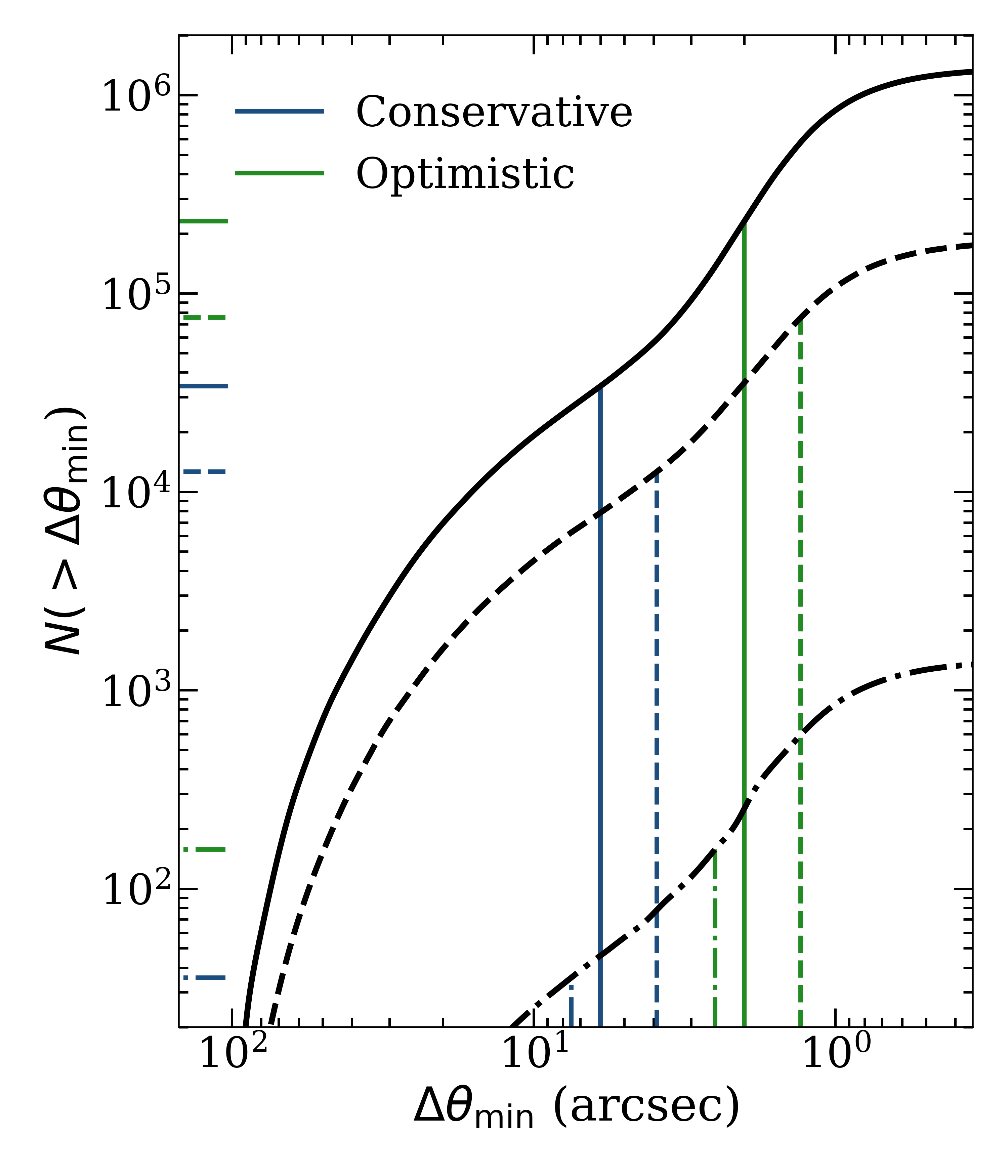}
    \caption{The cumulative total number of discoverable lenses in a DSA-2000 all-sky survey (solid), SKA-Mid AA* all-sky survey (dashed), and the VLASS (dash-dotted) as a function of the minimum angular separation between lensed images for discovery. The values of $\Delta\theta_\text{min}$ corresponding to our conservative and optimistic estimates are indicated in blue and red, respectively.}
    \label{fig:cumsep}
\end{figure}

\begin{figure}
	\includegraphics[width=\columnwidth]{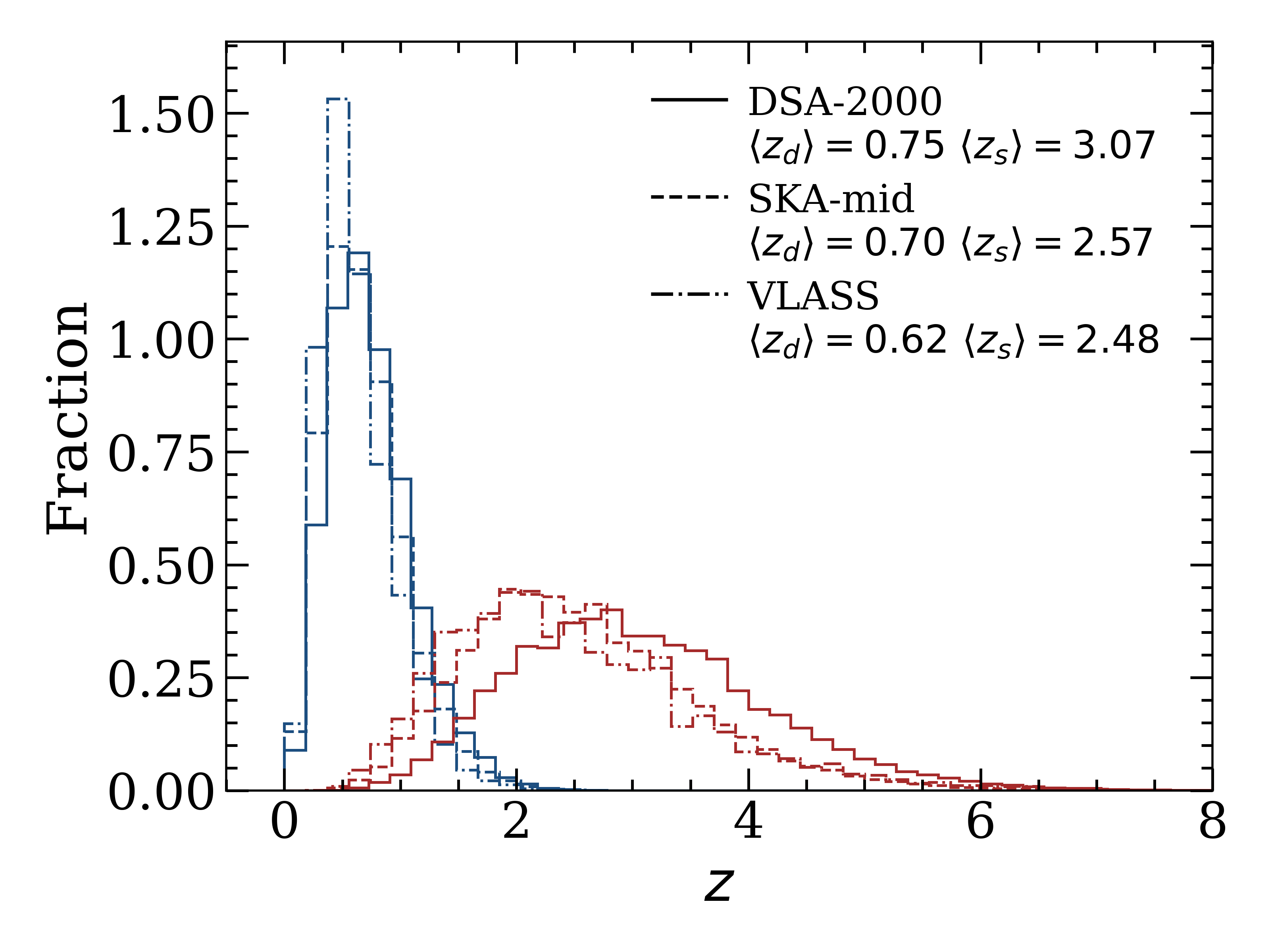}
    \caption{Redshift distribution of the deflectors (blue) and sources (red) in our optimistic prediction for the DSA-2000 (solid), SKA-Mid AA* (dashed), and VLASS (dotted)  with magnification bias. The legend indicates the average redshift of the deflectors, $\langle z_d\rangle$, and sources, $\langle z_s\rangle$.}
    \label{fig:sourcedeflector_dist}
\end{figure}

The final results of our simulation are reported in Tables \ref{tab:resultsbysource} and \ref{tab:resultsbydeflector}. Lensing yields for the DSA-2000 (optimistic) broadly agree with the empirical estimates given above
for the number of galaxy lenses and lensed SNe. The lensing optical depth and image separation distribution for galaxy-, group-, and cluster-scale lenses predicted by our model for the DSA-2000 are shown in Figures \ref{fig:tau} and \ref{fig:angular_sep} respectively. For the DSA-2000, SKA-Mid AA*, and VLASS, only a few percent of the total lenses above the flux-density limit meet the conservative image separation discovery limit. We expect most of the discoverable lenses in the conservative estimates for each survey to be group and cluster scale lenses. The large number of expected group and cluster scale lenses compared to known systems can be attributed to the very high sensitivities of the DSA-2000 and SKA-Mid. At $\mu$Jy flux-densities, the areal density of sources is such that one can expect roughly 0.5 background sources within the Einstein radius (a representative lensing scale) of every cluster \citep{Matthews_2021}. The total number of discoverable lenses in each survey as a function of  $\Delta\theta_\text{min}$ is shown in Figure \ref{fig:cumsep}. The difference in total lensing yields for each survey between the conservative and optimistic estimates is quite large. If lensing separation scales smaller than the conservative limit can be accessed by super-resolution or other techniques, galaxy-scale lenses will start to be identified, and the increase in total lenses will be significant. This motivates the development of these techniques for lens searches in the radio (see Section \ref{sec:discover}).
Figure \ref{fig:sourcedeflector_dist} shows the redshift distribution of the deflectors and sources. Almost all deflectors are at $z_d<3$ and most are at $z_d\approx1$, consistent with \cite{Oguri2010,Collett_2015,LensQSOsim,Yue_2022}. The average redshift of lensed sources is $\sim2.5-3$, with a small but significant number at very high redshift for the DSA-2000 and SKA-Mid. The expected percentage of systems with 4+ images in Table \ref{tab:resultsbysource} is large, especially for the conservative estimates, because group and cluster scale lenses tend to produce more multiple images than galaxy scale lenses.

We expect roughly $10^4-10^5$ lenses in each of the DSA-2000 and SKA-Mid all-sky surveys, and several hundred lenses in VLASS. This indicates at least a two-order-of-magnitude increase in the sample of radio lenses in the coming decade. As VLASS concludes its epoch 3, many of its lenses are likely already in the data (several unresolved lenses have recently been identified \citep{martinez2024findinglensedradiosources}). A lensing survey with the planned SKA-Mid AA4 array would have the advantage of being able to readily discover a large number of galaxy-scale lenses, significantly increasing its yield.
\cite{mckean2015ska} estimate $\sim3\times10^5$ galaxy-scale lenses in an SKA-Mid \text{AA4} all-sky survey with $\sigma_n=3\,\mu$Jy/beam and a lens detection limit of $15\sigma_n$ and >0.3''. This is slightly
more optimistic than our forecast, but broadly consistent. 

The expected number of galaxy-scale lenses in Rubin-LSST and the Euclid survey are $\sim$100,000 each \citep{Collett_2015,euclid2024,euclidcollaboration2025euclidquickdatarelease}. The DSA-2000 and SKA-Mid AA* will discover comparable numbers of galaxy-scale lenses if the PSF limitations can be overcome, while an SKA-Mid AA4 lensing survey would discover nearly as many even in the conservative estimate. \cite{LensQSOsim,Abe_2025} find that the Rubin-LSST will discover about 2000-3000 lensed QSOs. The number of lensed compact radio AGN (FRSQs and BL Lacs) expected for upcoming radio surveys is smaller but of the same order of magnitude. \cite{Oguri2010} predict that Rubin-LSST will find roughly 130 lensed SNe, while \cite{Abe_2025} predict as many as $\sim$200. We expect fewer lensed SNe at radio wavelengths for several reasons: not all SNe emit in the radio, the radio emission is weaker than in other wavelengths, and the cadence of radio surveys will be significantly longer than that of Rubin-LSST \citep{Bietenholz_2021}. 

\begin{figure*}
\includegraphics[width=\textwidth]{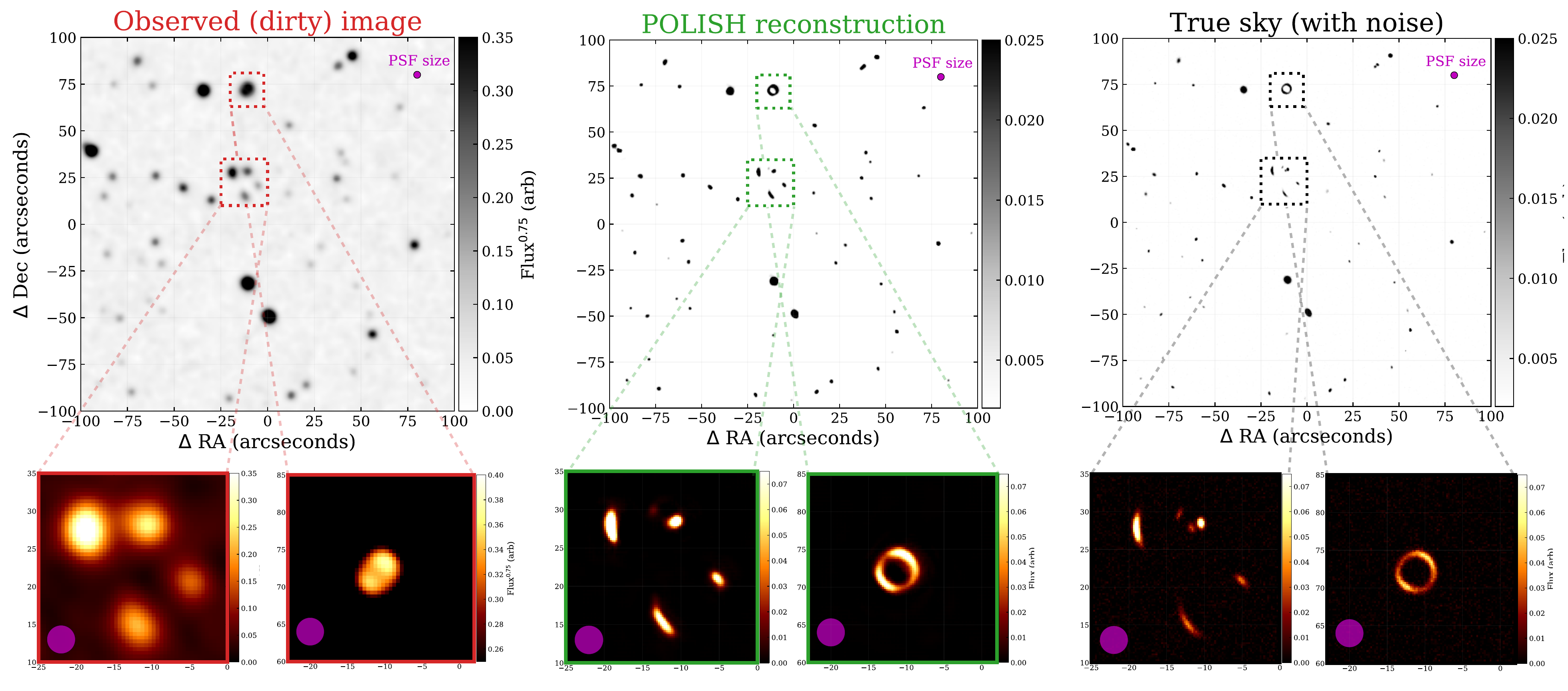}
    \caption{An example of super-resolution image reconstruction 
    on radio strong lenses. The simulated sky model is a mix of star-forming radio galaxies and AGN point sources. We randomly select 
    $5\%$ of sources to be strongly lensed. The artificially 
    high value ensures that we have several lensing examples
    to reconstruct. 
    The left panel shows 
    a 3.3'x3.3' region observed (i.e. the "dirty image") with the DSA-2000 full-band PSF (size shown with mauve circle) without any deconvolution. The 
    middle panel is a reconstruction of that field with the 
    POLISH algorithm. The rightmost panel is the true sky. The dirty images have been gamma encoded with $Flux^{0.75}$ with a value range of the inset figures chosen to highlight structure. In this toy example, strong lensing systems with Einstein radii 
    below the PSF scale can be identified.}
    \label{fig:polish}
\end{figure*}

\section{Lens discovery}
\label{sec:discover}

\subsection{Super-resolution}
\label{sec:sr}

In the past decade, tremendous progress has been made by the computer vision 
community with respect to the classical 
ill-posed inverse problems. These include 
deblurring \citep{zhang2022deep}, deconvolution 
and super-resolution \citep{superres}, and image inpainting \citep{yu2018generative}. 
Nearly all of these advances were borne out of the deep learning 
revolution, as efficient neural network architectures and 
training strategies have enabled powerful learning-based tools 
for machine vision.

Astronomy naturally lends itself to these tools because sparse sampling and ill-posedness arise in many astronomical imaging and reconstruction contexts, especially in interferometry. Several groups have begun developing machine learning methods 
for interferometric image reconstruction \citep{POLISH,r2d2,marspsf}. 
One reason this approach is suitable for radio astronomy 
is that the PSF of an array is given deterministically (barring calibration errors) by the 
spatial distribution of antennas and the observing frequency: 
The on-sky response of an interferometer is the 2D Fourier 
Transform of its sampling in UV-space (the aperture plane). 
Therefore, prior physical knowledge of the PSF can be 
incorporated in the model either via training 
data \citep{POLISH} or fed directly to the network itself \citep{marspsf}. This is not the case in ground-based 
optical astronomy where "seeing" and complex optics mean 
the PSF is not known a priori.

As a simple demonstration, we use the super-resolution and image-plane deconvolution method {\tt POLISH} to show how strong lenses 
with image separations near the PSF scale can be improved
with machine learning. {\tt POLISH} is a supervised machine learning model that learns the mapping between the true sky and observed images (the "dirty image" in radio parlance) \citep{POLISH}. It uses the Wide Activation for Efficient and Accurate Image Super-Resolution (WDSR) architecture \citep{wdsr}, but 
any super-resolution neural network could be swapped in (e.g., 
a standard U-Net or the Efficient Super-Resolution Transformer \citep{edsr}). 

We have trained a {\tt POLISH} network on forward-modelled 
synthetic data using 
a DSA-2000 PSF averaged over the whole band, resulting 
in an angular resolution of $\sim\,3.3''$. This is worse 
resolution than the top of the band, where many strong lenses could be found, but we offer this as a toy example. The sky model is described in \citet{POLISH}, with the addition of strongly lensed galaxies in both the training and validation set. In short, 
a realistic PSF was generated based on the antenna positions and observing frequencies of the DSA-2000. We then generated 
800 sky images with 2048 pixels per side. The sky model is populated 
with galaxies and AGN drawn from realistic brightness, size, and shape distributions. For our toy lensing example, we randomly select 1 in 20 galaxies to be strongly lensed. Noise is added to each image and it is convolved with the PSF after randomly perturbing the PSF 
to mimic calibration errors. A POLISH model was then trained on these image pairs, during which time the model learns a mapping between measurement and true sky. After training, the model can now deconvolve and super-resolve images observed with the same 
or similar PSF. At inference, the model is not explicitly given a PSF. It is learned from the training data implicitly. 

In Figure~\ref{fig:polish} we show an example validation image 
that contains multiple lenses. Lensed systems that are 
undetected in dirty image, including both 
arcs (left inset panels) and Einstein rings (right inset panels), are identifiable in the 
{\tt POLISH} reconstruction. 

As with any solution to an ill-posed inverse problem, it is important 
to carefully understand under what conditions POLISH fails and how results are impacted by one's prior (i.e. choice of regularization). This is especially true for an end-to-end machine learning model where a reconstruction prior is generated from the training data itself. It will therefore 
be essential to stress-test learning-based reconstruction algorithms with 
physically realistic forward models that include calibration errors (both pointing dependent and pointing independent) and physical lens models. Some of this has already been done by \citet{POLISH}, but under simplifying assumptions and without any gravitational lensing in the forward model. In the future, we plan to develop super-resolution methods explicitly for lens finding. The critical question that 
such simulations would answer is what fraction of lenses can be recovered and with 
what false-positive rates, as a function of lens type.
For now, we take the current demonstration as a promising sign that imaging techniques like POLISH could 
improve lens recovery on fill-aperture arrays such as DSA-2000. 

Finally, we point out that POLISH could be 
used on other instruments at different wavelengths, as long as two conditions are met: the PSF is known (e.g. space-based telescopes) and the sky and its 
noise properties can be accurately forward modeled. Such methods would therefore not be applicable to, say, Rubin, due to its being seeing limited. 
But perhaps Roman or Euclid would benefit from 
super-resolution techniques.

\subsection{Identification}

The first major hurdle for strong lensing science with DSA-2000 and other surveys is identification. The large number of sources produced by these surveys makes visual inspection by experts challenging. Despite much work on automated lens discovery (see \cite{lemon2023searchingstronggravitationallenses} for a review), little effort has been spent on these tools in the radio (the only such work to the authors' knowledge being \cite{Rezaei2022}). On the one hand, radio lens images are made cleaner by the lack of radio emission from 
the massive quiescent galaxies that make up the 
deflector population. However, the large, extended lobes of radio galaxies can easily be misidentified as lensing features. Further, color and photometric redshift information from UV and optical surveys have traditionally played a role in lens identification, despite being a foreground 
to the morphology of the lensed source.

Catalog-level searches, where large cuts are made based on certain features before a smaller subset is visually inspected, have been successful in the past \citep{CLASSI,Springola2018,Casadio2021}. CLASS selected for flat spectrum radio sources ($\alpha > -0.5$) \citep{CLASSI}. This had the advantage of picking out mostly compact radio sources, including blazars, thus eliminating any confusion with intrinsic structure. A similar strategy would likely be effective for the DSA-2000 or SKA. However, dedicated follow up of all of these sources, as in the $\sim$11,000 compact radio sources in the initial CLASS sample, will be observationally intractable. Similarly, a visual inspection of even a small subset of the $\mathcal{O}(10^7)$ blazars expected in the DSA-2000 is ambitious, hence the need for intermediate automated steps in the identification pipeline. Still, targeting flat-spectrum sources could increase the yield of any search. Spectral information from the whole DSA-2000 or SKA band will be useful for determining whether components are different sources or multiple images, and component surface brightness measurements can filter out core-jet sources \citep{Springola2018}. 

\citet{JacksonBrowne2006} show that combining astrometric data from radio and optical surveys can significantly improve the efficiency of lens searches. If the optical emission is dominated by the deflector galaxy then there will exist an offset between the centroid position of the optical and radio sources in a lens system. This offset was utilized to search for lenses with separation down to $\sim1''$ even with the poor $5''$ resolution of the Faint Images of the Radio Sky at Twenty-one centimeters (FIRST) survey. They predict that this effect will become more efficient at lower radio flux-densities because the optical flux-density is more likely to be dominated by the deflector galaxy.  In general, exploiting the extra information from the overlap between upcoming radio surveys and those at other wavelengths will be crucial for efficient radio lens searches, whether this is simple cross matching, e.g. \cite{martinez2024findinglensedradiosources}, or searching for new lenses. Many of the known radio lenses were selected in other wavelengths or found by combining radio data with other wavelengths \citep{Kayser1990,Ibata1999,Lacy2002,Carilli2003,Garrett2005,Haarsma2005,Inada_2006,Berciano2007,Ghosh2009,Ivison2010,McKean2011,Mckean2011b,Jackson2011,Valtchanov2011,Messias2014,Geach2015,Jackson2015,DessaugesZavadsky2017,Hartley2021,Mangat2021,Glikman2023,Giulietti_2023,Gross2023,Dux2023,Chen2024,jackson2024radio,dobie2024gaia}. Given the sensitivity and wide-area overlap of the DSA-2000's all-sky survey (and SKA-Mid if it undertakes a wide-field survey) with next-generations O/IR surveys, and the $\sim$10$^5$ lenses expected in those surveys \citep{Collett_2015}, these methods are promising for radio lens finding and overcoming the limitations of the DSA-2000's PSF.
For example, cross-matching Rubin-LSST discovered lenses with DSA-2000 or SKA-Mid data should yield $\sim$thousands of new radio lenses with relative ease. 

Machine learning models have been successfully used to discover many new candidate lens systems \citep{lemon2023searchingstronggravitationallenses}. CNNs in particular are effective at classifying astronomical images.  Using CNNs on simulated data for the ILT at 150 MHz, \cite{Rezaei2022} are able to recover over 90\% of galaxy-size lenses with a false-positive rate of only 0.008\%. They find that a  $20\sigma$ detection and image separation $\theta_E\geq3/2$ beam size are necessary for reliable identification with the CNN. Incorporating super-resolution models into the lens-finding routine could significantly increase the yield of such a CNN, especially at high SNR. Super-resolution methodologies have not been validated for lens finding, but show promise for enabling the discovery of lenses below the PSF limit (Section \ref{sec:sr}). \texttt{POLISH} can recover scales down to $\sim$1'' for the DSA-2000 \citep{POLISH}, which could enable resolving lenses at or below the PSF scale at the same level of accuracy as in \cite{Rezaei2022}. The main disadvantage with CNNs and other supervised learners is the need for large realistic training sets \citep{lemon2023searchingstronggravitationallenses}. \cite{Rezaei2022} use simple Gaussian components to model background radio sources. In reality, extended radio sources can have complex morphologies. Additional complications are added by image processing artifacts. Given that even the best O/IR automated lens finders have a purity of $\lesssim$10\% when applied to real data (e.g. \cite{PearceCasey2025,euclidcollaboration2025euclidquickdatarelease}), effective radio lens finders will require advanced radio source and instrument response models. For the DSA-2000, its filled aperture array layout will produce very clean images with minimal beam sidelobes and other image artifacts \citep{hallinan2019dsa2000radiosurvey}, especially compared to the ILT, meaning that a CNN trained specifically for the DSA-2000 could do better at lens finding than the one in \cite{Rezaei2022}. Simulating accurate DSA-2000 lenses and training models for identifying them is a goal of our future work. Given the rapidly developing field of ML, it is also probable that significant progress will be made on automated lens finders by the end of the DSA-2000's all-sky survey in about a decade; for example with new architectures like Vision Transformers \citep{gonzalez2025discoveringstronggravitationallenses}. The SKA-Mid AA4 array will already have sufficient resolution to find the majority of galaxy-scale lenses without super-resolution or other additions to the lens finder discussed in this section, making CNNs an attractive option for lens identification if it undertakes an all-sky survey. The AA* array, with its intermediate resolution, will not require additions beyond current CNNs to find galaxy-scale lenses, but its yield could be increased significantly with them.

Group and cluster lens identification is nearly uncharted territory in the radio. Known large separation radio lenses were all selected in other wavelengths \citep{Garrett2005,Inada_2006,Berciano2007,Ghosh2009,McKean2011,Jackson2011,DessaugesZavadsky2017}. An early catalogue-level search for $6-15''$ separation lenses in the CLASS sample was unsuccessful \citep{Philips2001}, as were similar searches up to $60''$ in FIRST data \citep{Philips2001b,Ofek2001}. In the O/IR, group and cluster lenses are typically found by examining the fields of known high-mass systems \citep{Gladders_2003,Jaelani2020}, with automated searches for arc-like features \citep{More_2012}, or with catalogue-level searches \citep{Belokurov2008,Inada_2008}. It is not clear whether radio cluster lenses will show the same clear arc features as in the O/IR (e.g. \cite{Garrett2005,Berciano2007}). However, because 1) the probability of lensing for each of these high mass systems is relatively large, 2) we expect at the sensitivity of the DSA-2000 or SKA-Mid the majority of these systems will have background sources within $\sim\theta_\text{E}$, and 3) the total number of these systems is small compared to galaxies, visual examination of large catalogues of group or cluster systems to identify the majority of the expected large-separation lenses is not intractable. The main limitation of any learning-based approach to group or cluster lens finding is the difficulty of creating realistic training sets. The irregularity of group and cluster lensing potentials may require building a forward-modelled training set from ray-tracing 
in large cosmological magnetohydrodynamical simulations such as TNG-Cluster, which should have a heterogenous population of massive clusters \citep{tngcluster}. More generally, a realistic training set 
of strong lenses across scales could be produced with radio source 
samples from T-RECS and a deflector population drawn from cosmological simulations. It should be noted that group and cluster lens identification will be complicated by the radio emission of cluster members, as seen in \cite{McKean2021,Heywood_2021}.

\section{Strong lensing applications}

A key limitation of strong lensing science is the small number of known systems, especially at radio wavelengths. 
We discuss the impact that the large expected number of radio lenses and their multi-wavelength counterparts will have on several important applications of strong lensing.

\subsection{Time-delay cosmography \& $H_0$}
\label{sec:h0}
Light from multiple images in a gravitational lens will reach the observer at different times. For continuum sources, we do not observe the difference in arrival times, but transients or time-variable sources allow us to measure the time-delay.
This time-delay is due to a difference in path lengths and gravitational time dilation near the deflector. As such, the time-delay encodes information about the geometry of the universe, and is inversely proportional to $H_0$ \citep{refsdal}. Much work has been dedicated to measuring these time-delays and using them to constrain $H_0$ \citep{Treu_2016, Birrer2024}. The H0LiCOW project has measured $H_0$ to 2.4\%, which is independent of and competitive with state-of-the-art $H_0$ constraints \citep{Wong_2019}. A large increase in the number of measurable time-delay systems will allow even tighter constraints and could help settle the current Hubble tension. 

We estimate the number of lensed variable AGN that will be useful for measuring $H_0$ in the DSA-2000 and SKA-Mid all-sky surveys. To first order, the variable AGN will be the blazars (FSRQ and BLLac type sources), as many as several thousand of which are predicted to be lensed and discoverable in the DSA-2000 and SKA-Mid data by our simulation. In principle, we expect many radio-selected blazars to be bright \citep{Mao_2016} and variable \citep{Zhang_2017,Abrahamyan2019} in other wavelengths. This means that a potentially large fraction of these lenses will be good candidates for dedicated time-delay studies in the O/IR. However, we focus here on how many of these systems will be useful for time-delay studies in the radio.

There are several known radio time-delays, coming mostly from  the CLASS lenses: B1608+656 \citep{Fassnacht1999,Fassnacht_2002b}, B1422+231 \citep{Patnaik2001}, B0218+357 \citep{Biggs1999,Cohen2000}, B1030+074 \citep{Biggs2018}, B1600+434 \citep{Koopmans2000}, 0957+561 \citep{Haarsma1999}, and PKS 1830-211 \citep{vanOmmen1995,Lovell1998}. The data were taken over two decades ago for all of these lenses, although there have been some recent improved analyses \citep{Biggs2018b,Biggs2018,Biggs2021,Biggs2023}. B1608+656 is one of the H0LiCOW lenses \citep{Wong_2019}, using the time-delay determined by \cite{Fassnacht1999,Fassnacht_2002b}. The lack of more recent time-delay measurements in the radio is due to the small number of known radio lenses and the limited availability of interferometer observing time for long monitoring campaigns. 

Several factors complicate our ability to use lensed blazars for time-delay measurements; we would like to know how many will be in a "gold sample" for which the time-delay can be reliably determined.
The quality of a time-delay measurement depends on the cadence and duration of observations, the photometric errors, the intrinsic variability of the source, and extrinsic variability (e.g. by microlensing) \citep{Birrer2024}. Estimating the effect of the individual photometric uncertainties of the data points is challenging because they propagate to the final time-delay in a non-trivial way that depends on the method used to extract the time-delay. However, in general, percent-level uncertainties on time-delay measurements are desired, so percent- or sub-percent-level photometric errors are necessary. The error in a flux-density or polarized flux-density measurement is a combination of the rms thermal noise and systematic error, such as calibration error. The systematic error has typically been comparable or larger than the thermal error in previous radio time-delay measurements at the level of $\lesssim1\%$ \citep{Biggs1999,Koopmans2000,Patnaik2001,Fassnacht_2002b,Biggs2018}. Predicting the systematic/calibration error for future radio telescopes is beyond the scope of this paper, but here we can assume that advances in radio astronomy since the early 2000s ensure the errors are still $\lesssim1\%$. The DSA-2000 in particular will be extremely calibratable due to its antenna layout (see the discussion in \cite{Byrne2024}). So it suffices to ensure that the rms noise is also at the percent-level in order to obtain time-delay measurements with comparable accuracy to the known radio time-delays. For example, the observations for the time-delay measurement of B1608+656 (which is accurate enough to be included in the H0LiCOW analysis), have rms noise $\sim0.1$mJy, yielding a SNR ranging from a $\sim$200 to $\sim$40 for the different components. We set $S_\text{min}=0.2$\,mJy for the lensed blazars in our gold sample, so that the total SNR for 1\,hr integrations with the DSA-2000 or SKA-Mid will be $\sim$100, and $\sim$1000 for the ngVLA. 

We also select for lensed blazars with $\geq10\%$ intrinsic variability. It is shown in \cite{Fassnacht_2002b} that an increase in intrinsic variability of the source from 5\% to $\sim$25-30\% allowed a factor of 2-3 better determination of the time-delay. 
To date, the most extensive study of radio AGN variability is the OVRO 40\,m blazar monitoring campaign \citep{ovro2011,Richards2014}. Since 2007 this program has monitored around 1500 blazars with a cadence of two weeks. \cite{Richards2014} provides distributions of the intrinsic modulation index, a measure of the relative intrinsic variability of the AGN defined as $m=\sigma_S/\langle S \rangle$, where $\sigma_S$ is the standard deviation of the flux-density and $\langle S \rangle$ is the average flux-density, for both FSRQs and BL Lacs. 
We use the statistics of the radio-selected CGRaBS sample from \citep{Richards2014}. The mean flux-density of the blazars at 15\,GHz is $\gtrsim 60$\,mJy, which is not representative of the majority of blazars we expect from the DSA-2000 or SKA-Mid \citep{ovro2011}. However, because no correlation between flux-density and modulation index is found for $S\geq0.4$\,Jy \citep{ovro2011}, and no statistical study exists for the faint blazar population, we assume that the statistics of the OVRO sample apply to all blazars. Further, because a marginal ($<2\sigma$) negative correlation between FSRQ modulation index and redshift is found \citep{Richards2014}, we use the statistics of the sample of "high" redshift ($z>1$) FSRQs, which is more representative of the lensed FSRQs we are considering. The OVRO blazars are monitored at 15\,GHz but we expect lower variability at 1.4\,GHz. We assume the mean modulation index drops by 40\% from the OVRO 40\,m data to 1.4\,GHz for the DSA-2000 and SKA-Mid \citep{Fan2006agnvariability,Sotnikova_2024}.

So far we have only considered lensed AGN for time-delay measurements, which are desirable because of their characteristic variability and compact emission regions. Lensed transients will also be useful for time-delay measurements, provided that they are detected early and monitored closely. Several lensed type 1a SNe have been detected at other wavelengths, some of which may have measurable time-delays \citep{Kelly2015,Goobar2017,Rodney2021}. The variability of radio emission from TDEs is too slow to make them useful for this application (on-axis jetted TDEs can have shorter rise times, but they are significantly rarer, such that we do not expect any lensed jetted TDEs in upcoming radio surveys), but radio ccSNe have faster rise times \citep{Alexander2020,Bietenholz_2021, cendes2023ubiquitouslateradioemission}. We retain the same discoverability limits from Section \ref{sec:dlimit} for considering lensed SNe for time-delay measurements. 

\begin{table}
    \centering
    \begin{tabular}{l@{\hspace{.7cm}}c@{\hspace{.7cm}}c}
        \hline
        Survey & Blazars & ccSNe \\
        \hline
        \hline
        DSA-2000  & 11 (45\%),\: 78 (19\%) & 17 (48\%),\: 64 (44\%) \\
        SKA-Mid AA* & 18 (41\%),\: 205 (11\%) & 4 (51\%),\: 18 (45\%) \\
        SKA-Mid AA4 & 243 (11\%),\: 520 (9\%) & 31 (40\%),\: 70 (37\%) \\
        \hline
    \end{tabular}
    \caption{The predicted number of lensed time-variable radio blazars (FSRQ and BL Lac) in a "gold sample" ($S_\text{min}\geq0.2$\,mJy and $m\geq0.1$) and radio ccSNe that would be useful for time-delay $H_0$ studies for the DSA-2000 and SKA-Mid all sky surveys. Each cell contains two values corresponding to our conservative and optimistic estimates, in that order, as defined in Section \ref{sec:dlimit}. The percentages indicate the systems with 4+ multiple images.}
    \label{tab:goldsample}
\end{table}

The number of discoverable lensed time-variable AGN in our gold sample for the DSA-2000 and SKA-Mid, based on our criteria of $S_\text{min}=0.2$\,mJy and $m\geq0.1$, is shown in Table \ref{tab:goldsample}, alongside the number of expected lensed ccSNe. Of the lensed AGN, the most useful systems for time-delay analysis will be those that have four or more images because they allow for multiple constraints on the time-delay, so we determine the number of these systems as well. We note that quad systems often 
have higher external shear due to their preponderance of 
group or cluster environments, which results in an added 
systematic in lens modeling \citep{Holder_2003}. 

\cite{Wong_2019} reach a 2.4\% measurement of $H_0$ with a sample of 6 lensed quasars, and predict that around 40 lenses are needed to constrain $H_0$ to within 1\%. \cite{Birrer2024} also estimate that a sample of ~40 lenses are needed for a 1\% measurement. \cite{Napier_2023} reach a 10\% measurement with three galaxy clusters, and estimate that roughly 50 will be needed for a sub 1\% constraint, assuming that modeling uncertainties can be reduced in coming years. If we assume that all of the lenses in our gold sample can be modeled to the accuracy of the H0LiCOW sample -- a roughly 6\% combined uncertainty on the time-delay, mass model, and line of sight contribution for each lens -- then a percent-level determination of $H_0$ with radio lenses should be feasible, even for our conservative estimate, by combining samples from both the DSA-2000 and SKA-Mid and known time-delay lenses. If separations near the PSF scales (the optimistic estimate) can be recovered, our forecast indicates that a sub-percent-level measurement may even be possible in the coming decade. This is of course not possible with the DSA-2000 or SKA-Mid alone but will require dedicated follow-up at multiple wavelengths to measure redshifts, densely sample  light curves, and constrain lens models.

It is worth noting that radio-selected lenses will have certain advantages for time-delay studies; for example, extremely high-resolution interferometry can enable very precise lens models (e.g. \cite{Spingola2019}, \cite{Powell2021}, and \cite{Stacey2024}), lensed compact radio sources may be less susceptible to microlensing \citep{Birrer2024}, and polarization information can aid in time-delay determinations \citep{Biggs1999,Biggs2018b,Biggs2018,Biggs2021,Biggs2023}. However, the main criteria will be the cadence and duration of observations.
Considering the possibility of measuring radio time-delays even without external follow-up, we note that special systems could be visited 
with a higher cadence by the DSA-2000, similar to how pulsar fields will be visited regularly for timing experiments. Both the SKA-Mid and the ngVLA will have the resolution necessary to accurately model lenses and the cadence to measure time-delays. The DSA-2000, which has a set plan for an all-sky survey and is optimized for survey speed, will discover a large number of lenses that can then be followed up with these higher-resolution arrays. 

\subsection{Dark Matter}

Gravitational lensing has been very useful for determining the (sub)structure of galaxies, groups, and clusters at cosmological distances and distinguishing between dark matter models (see \cite{Vegetti2024, Natarajan2024} for a review). 
Long wavelength observations offer a significant advantage for dark matter studies because of the extremely high angular resolution that can be achieved with interferometry -- important for lens modeling.
Further, VLBI will enable the detection of lens perturbations by very small subhalos, e.g. a 10$^6M_\odot$ subhalo is expected to alter image positions on the scale of milliarcseconds \citep{Vegetti2024}. 
Because resolution is a key factor in the ability to accurately model mass distributions or detect substructure, the DSA-2000 or SKA-Mid AA* alone will not be able to use lens systems to study dark matter. A fruitful strategy will be identifying large numbers of lens systems in DSA-2000 or SKA-Mid all-sky surveys, and then obtaining detailed follow-up to search for sub-halos. For unresolved sources, the high sensitivity of the DSA-2000, SKA-Mid, or ngVLA will enable precise determinations of flux-ratios. For resolved sources, an ngVLA or VLBI survey of the expected large sample of radio lenses will likely uncover many milliarcsecond scale perturbations and revolutionize this field. 

Another exciting application of lensing is using a central image to study small scales near the centers of galaxies, such as the central supermassive black hole \citep{Mao2001,Rusin_2005,Treu2010,mckean2015ska, shajib2024stronglensinggalaxies}. 
This is especially suited for radio wavelengths because in many cases the deflector will be radio-quiet, allowing the center image to be detected. Because the center image is highly demagnified, only sensitive radio telescopes such as the DSA-2000 or SKA-Mid will be able to reliably identify them. One such central image has been detected in the radio with flux-density $\sim 0.8$\,mJy at 8.46\,GHz \citep{Winn2004}.
From our simulation, we expect, for the DSA-2000, about 0/500 galaxy-scale lens sytems with an odd number of images including a central image (minimum $|\pmb{u}|$) that is demagnified ($|\mu|<1$) but still with flux-density above $5\sigma_n$, for the conservative/optimistic estimates. For the SKA-Mid AA* and AA4 arrays, this is 0/400 and 700/2800, respectively. This agrees with the $\sim10^{-4}-10^{-5}$ estimate for the rarity of these images given by \cite{mckean2015ska}. A main challenge will be disentangling the faint emission of the central image from the other images. Further, at $\mu$Jy flux-densities, the faint radio emission of lensing galaxies may become an issue in identifying central images.

\subsection{Other applications}

An advantage of radio surveys is measuring polarization information, which is conserved under lensing. This enables 
us to study propagation effects along different 
sightlines for lensed objects \citep{Greenfield85}. For example, \citet{Mao_2017} used the VLA to measure the polarization properties of two lensed images of CLASS B1152+199, 
finding a large difference in RM 
which is due to the magnetized plasma 
in the lens galaxy's interstellar medium. 
The DSA-2000 will have the ability to 
measure full polarization information for 
any source it detects, with a maximum |RM| 
of roughly $10^4$\,rad\,m$^{-2}$. Across all radio AGN, typical polarization fractions at these frequencies are a few percent (or more for compact sources) \citep{farnes2014}.
Assuming a typical value of 5$\%$ \citep{beckandgaensler} and  a 100\,$\sigma$ detection 
threshold in total power (i.e. a polarization detection 
of $\geq5\,\sigma$), we 
estimate that the DSA-2000 could detect 
Stokes Q and U for 3-5 million AGN in its 5 year continuum survey. 
Using the empirical CLASS lensing optical depth 
and image separation cut, we take
$\tau_{AGN}\approx5\times10^{-4}$.
We estimate that the DSA-2000 could find 
$\mathcal{O}(10^3)$ strong lenses for which 
polarization properties could be used to 
model the lens distribution and 
study magnetic fields at cosmological 
distances.

If multiple sources at different redshifts are lensed by the same deflector, e.g. the "Jackpot" lens \citep{gavazzi2008sloan}, the ratio of the angular diameter distances to the sources can be determined and thus cosmological parameters such as the dark energy equation of state $w$ and $\Omega_m$ \citep{Birrer2024}. 
A large influx in observed cluster lenses with the DSA-2000 or SKA-Mid will allow better constraints on these cosmological parameters. In a very rough estimate, for $\sim5\times10^8$ DSA-2000 sources we expect that about half of all clusters will have a background radio source within their Einstein radius. Assuming a Poisson distribution, about one in every 20 clusters should have two sources, meaning that we can likely expect $\mathcal{O}(10^3)$ double source plane cluster lenses with the DSA-2000.

Lenses are also commonly used as "nature's telescopes", enabling the study of extremely small and faint objects. 
We expect to see $\sim$2700/11000 lenses with $z_s>5$ in the DSA-2000 for the conservative/optimistic estimates, and $\sim$340/1300 for the SKA-Mid AA* (Figure \ref{fig:sourcedeflector_dist}), many of which will be group and cluster lenses due to their high magnifications. These high-redshift sources will provide insight into the population of radio sources in the early universe.

\section{Conclusions}
In this paper, we forecast expected strong lensing yields in the upcoming DSA-2000 and SKA-Mid wide-field radio surveys, as well as the current VLASS. We develop a forward model that accounts for different deflector and radio source populations and current expectations for the performance of these instruments. Notably, we model the expected number of galaxy group- and cluster-scale lenses because these systems will be easily discovered due to their wide angular separations. We find that both the DSA-2000 and the SKA-Mid will discover roughly $10^4-10^5$ strong lens systems, depending on the minimum angular separtation of lensed images that can be recovered, while there should be as many as $\sim100$ lenses already contained in VLASS data. We discuss strategies for identifying these lenses, which will all likely benefit from emerging super-resolution 
techniques. Finally, we discuss the scientific application of the huge numbers of lenses that will be discovered by these surveys. One of the most exciting applications is $H_0$ cosmography with variable and transient sources. The DSA-2000 and SKA-Mid will each discover roughly $10^1-10^2$ lensed flat spectrum AGN that will be good candidates for radio time-delay measurements, as well as $\sim20-70$ lensed radio ccSNe. With dedicated multi-wavelength follow up these systems could be used to constrain $H_0$ to $\sim$1\%. The new lens systems will also be useful for studying the distribution of dark matter at cosmological distances, among other applications.

\section*{Acknowledgements}
We are grateful to Schmidt Sciences for supporting 
Samuel McCarty as a Summer Undergraduate Research 
Fellow at Caltech.
We thank Kim-Vy Tran, Tony Readhead, and 
Tommaso Treu for helpful conversations on strong lensing, as 
well as Paul Schechter for insights into quad systems.

\section*{Data Availability}
The code used to produce this paper is available on the public GitHub repository at  \url{https://github.com/smmccrty/radiolensing_pub}.



\bibliographystyle{mnras}
\bibliography{example}


\bsp
\label{lastpage}

\end{document}